\documentclass{mn2e}

\usepackage{natbib-MS,graphics,epsfig}

\def\spose#1{\hbox to 0pt{#1\hss}}
\def\lta{\mathrel{\spose{\lower 3pt\hbox{$\mathchar"218$}}
     \raise 2.0pt\hbox{$\mathchar"13C$}}}
\def\gta{\mathrel{\spose{\lower 3pt\hbox{$\mathchar"218$}}
     \raise 2.0pt\hbox{$\mathchar"13E$}}}

\def\1p{\phantom{0}}

\newcommand\apj{{ApJ}}%
\newcommand\apjl{{ApJ}}%
\newcommand\apjs{{ApJS}}%
\newcommand\aap{{A\&A}}%
\newcommand\mnras{{MNRAS}}%
\newcommand\solphys{{Sol.~Phys.}}%
\newcommand\ssr{{Space~Sci.~Rev.}}%
\newcommand\nat{{Nature}}%
\title[Solar magnetic flux tube simulations]
{Solar magnetic flux tube simulations with time-dependent ionization}
\author[D. E. Fawzy, M. Cuntz, W. Rammacher]
{D. E. Fawzy$^1$\thanks{E-mail: diaa.gadelmavla@izmirekonomi.edu.tr (DEF);
cuntz@uta.edu (MC); wrammacher@online.de (WR)}, M. Cuntz$^2$ and W. Rammacher$^3$ \\
$^{1}$Faculty of Engineering and Computer Sciences, Izmir University of Economics, Izmir 35330, Turkey \\
$^{2}$Department of Physics, University of Texas at Arlington, Box 19059,
Arlington, TX 76019, USA \\
$^{3}$Kiepenheuer-Institut f\"ur Sonnenphysik, Sch\"oneckstr. 6, 79104 Freiburg, Germany
}

\begin{document}

\date{Accepted 2010 .....; Received 18.07. 2010}

\pagerange{\pageref{firstpage}--\pageref{lastpage}} \pubyear{2009}

\maketitle

\begin{abstract}
 In the present work we expand the study of time-dependent ionization
 previously identified to be of pivotal importance for acoustic waves in
 solar magnetic flux tube simulations.  We focus on longitudinal tube waves (LTW)
 known to be an important heating agent of solar magnetic regions.
 Our models also consider new results of wave energy generation as well as an
 updated determination of the mixing length of convection now identified as
 1.8 scale heights in the upper solar convective layers.  We present 1-D wave simulations
 for the solar chromosphere by studying tubes of different spreading as function
 of height aimed at representing tubes in environments of different magnetic filling
 factors.  Multi-level radiative transfer has been applied to correctly represent
 the total chromospheric emission function.  The effects of time-dependent ionization
 are significant in all models studied.  They are most pronounced behind strong
 shocks and in low density regions, i.e., the middle and high chromosphere.
 Concerning our models of different tube spreading, we attained pronounced
 differences between the various types of models, which were largely initiated by different
 degrees of dilution of the wave energy flux as well as the density structure
 partially shaped by strong shocks, if existing.  Models showing a quasi-steady rise
 of temperature with height are obtained via monochromatic waves akin to
 previous acoustic simulations.  However, longitudinal flux tube waves are
 identified as insufficient to heat the solar transition region and corona in
 agreement with previous studies.
\end{abstract}



\section{Introduction}

The outer atmospheric activity of the Sun as well as other late-type stars
is largely determined by the structure and time evolution of photospheric
magnetic fields.  These fields extend into the stellar outer atmosphere,
where they cause nonradiative energy to be deposited
\citep[e.g.,][]{and89,lin91,schr01,bog03,rob04,ulm04,mus04}.
This energy is distributed over the chromosphere, transition region, and
corona; it is also considered pivotal for the heating and acceleration of
solar and stellar winds.  For new observations and simulations of
the relative importance of acoustic and magnetic energy deposition in
the solar chromosphere, see, e.g., \cite{fos05}, \citet*{cun07}, \cite{kal07},
and \cite{bel09}.  Recent joint observational and theoretical studies
aimed at elucidating the significance of wave heating in the solar
photosphere and lower chromosphere were given by \cite{bec10},
\citet*{bec12}, and \cite{wed12}.  The latter study investigates the
channeling of magnetic energy through the solar chromosphere into the
corona, thus resembling super-tornadoes under solar conditions.
Notable reviews aimed at solar chromospheric and coronal heating were
given by \cite{nar96}, \cite{kli06}, and \cite{erd07}.

Magnetic flux tubes are an important feature of the solar surface
structure \citep[e.g.,][]{spr83,sol96,schr00}.  Both observational and
theoretical studies showed that they are carriers of longitudinal tube waves
\cite[e.g.,][]{her85,sol92,has08},
which give rise to considerable temperature increases as function of
height as revealed by chromospheric spectral features.
In this paper, we continue to pursue our previous line of research
focused on the generation of different types of waves
\citep[e.g.,][]{nar96}, effects of the propagation and
dissipation of waves \citep*[e.g.,][]{her85,faw98,cun99}, and the
emergence of chromospheric emission \citep[e.g.,][]{cun99,faw02,ramc03}.
However, in accord with previous simulations of acoustic waves
\citep[e.g.,][]{car92,car95,ramu03,ramc05b,cun07}, the described
longitudinal wave simulations will also employ time-dependent
(i.e., non-instantaneous) ionization.  The geometrical and thermodynamic
properties of the tube atmospheres are expected to also impact the
dissipation of the wave energy as well as the so-called energy velocity
as recently pointed out by \cite{wor12}.  Time-dependent ionization
entails that, e.g., behind strong shocks the long timescales
of hydrogen ionization / recombination initially prevent the dissipated
energy to be converted into ionization energy, thus leading to strong
temerature spikes as well as a variety of other dynamic phenomena.

There exists a great motivation to revisit the dissipation of
acoustic and magnetic waves in the solar chromosphere, which is
the new determination of the mixing length near the top of the
solar convective zones.  \cite{stei09a,stei09b} who provided
new state-of-the-art simulations of the solar convection zone
on the scale of supergranules, extending 10\% of its depth but
half of its pressure scale height, deduced a mass mixing length
of $\alpha_{\rm ML} = 1.8$, thus superseding the previous results
of 1.5 or 2.0 \citep{stef93,tra97}, which were widely used in
previous solar heating computations.

In our study we will also consider the relevance of tube spreading,
i.e., different tube opening radii, for the energetics and
thermodynamics of the magnetically heated solar chromospheric
structure.  Early results based on adiabatic longitudinal waves
without the consideration of time-dependent ionization have been
given by \cite{faw98}.  They found that the tube shape is of
critical importance for the heating of flux tubes.  In fact,
tubes of wide opening radii show little heating, whereas constant
cross section tubes show very large heating at all heights.
Previous simulations of longitudinal waves for solar coronal
tube structure have been given by, e.g., \citet*{ofm99},
\citet*{ofm00}, and \cite{cuns04}.
These results further highlight the importance
of tube spreading with respect to the shock wave amplitude and
time-dependent and height-dependent heating rate.  Concerning
the formation of Ca~II and Mg~II emission, we will consider a
two-component (magnetic and acoustic) model of the solar
chromosphere with heating by longitudinal tube waves inside the
flux tubes and heating by acoustic waves outside the flux tubes.

The governing equations as well as the methods of our study,
including the computation of the initial flux tube models,
are discussed in Sect.~2.  In Sect.~3, we describe the results
of our longtitudinal wave simulations for solar flux tube models
with different tube spreadings; the latter correspond to
different magnetic filling factors.  The focus of these studies
concerns the effects of time-dependent ionization.  Our summary
and conclusions are given in Sect.~4.


\section{Methods}

\subsection{Theoretical approach}

In the following, we summarize some of the key equations with focus on
longitudinal MHD tube waves propagating along the vertically directed
magnetic flux tubes.  Following previous work by, e.g., \cite{her85},
the equations to be solved in the Euler frame for continuous 1-D flows are
given as
\begin{equation}
\frac{\partial \rho A}{\partial t}+\frac{\partial \rho uA}{\partial x} \ = \ 0
\ ,
\label{e21}
\end{equation}
\begin{equation}
\frac{\partial u}{\partial t}+u\frac{\partial u}{\partial x}
+\frac{1}{\rho}\frac{\partial p}{\partial x}+g\left( x\right) \ = \ 0
\ ,
\label{e22}
\end{equation}
\begin{equation}
\frac{\partial S}{\partial t} + u \frac{\partial S}{\partial x} \ = \
\frac{dS}{dt} \Big|_{{\rm rad}} \ ,
\label{e23}
\end{equation}
\begin{equation}
\Phi \ = \ A\cdot B \ = \ {\rm const}
\ ,
\label{e24}
\end{equation}
\begin{equation}
{B^2 \over {8\pi}} + p \ = \ p_e \ = \ \epsilon p
\ .
\label{e25}
\end{equation}
Here all variables have their usual meaning.  Note that
$p$ and $p_e$ are the gas pressures inside and outside the
tube, respectively, $B$ is the magnetic field strength, and
$\epsilon = p_e : p$ denotes the pressure ratio
between outside and inside of the tube.
Furthermore, the cross section $A$ is time-dependent owing
to the distensibility of the tube \citep[e.g.,][]{lig78}.
Moreover, Eqs.~(\ref{e24}) and (\ref{e25}) represent the
conservation of the magnetic flux $\Phi$ and the horizontal
pressure balance between inside and outside the tube.  All
quantities are both functions of height $x$ and time $t$;
strictly speaking, this also applies to $p_e$ where we
however assume $p_e=p_e(x)$ in accord to previous studies
\cite[e.g.,][]{her85,cun99}, as we do not consider the
effects of the time-dependency of the outside medium
toward the tube.

Similar to the case of acoustic waves \cite[e.g.,][]{ulm77,cun88},
the system of equations can be transformed into its characteristic
form as pursued in detail by \cite{ramu03}.  A key equation concerns
the behaviour of the thermodynamic variables along the C$^0$
characteristics with the relevant compatibility relation given as
\begin{equation}
dS = \frac{dS}{dt} \Big|_{{\rm rad}} dt
\ ,
\label{e07}
\end{equation}
which can also be written as
\begin{equation}
dT - \left( \frac{\partial T}{\partial p} \right)_S dp -
\left( \frac{\partial T}{\partial S} \right)_p
\frac{dS}{dt} \Big|_{{\rm rad}} dt = 0
\ .
\label{e09}
\end{equation}
Note that, e.g., $\left( \partial T / \partial p \right)_S$
and $\left( \partial T / \partial S \right)_p$ constitute
well-known thermodynamic relationships \citep[e.g.,][]{mih84};
they need to be solved numerically if deviations from
the thermodynamic equilibrium, as in case of our study,
are considered.

The rate equations for the general case of flows have been
discussed in detail by, e.g., \cite{mih84} and have
been implemented into actual OHD/MHD codes by \cite{car02},
\cite{ramu03}, among others.  Assuming a simplified hydrogen
atom of $N=3$ levels, namely two bound levels plus the
continuum level, while considering radiative processes $R$
and collisional processes $C$, the population density of
level $i$ can be obtained through solving
\begin{equation}
{\partial n _i\over \partial t} +{\partial  n_i u\over \partial x} \ = \
\sum_{j = 1~\land~j \neq i }^{N} \left( n_j P_{ji} - n_i P_{ij} \right)
\ ,
\label{e54}
\end{equation}
where $P_{lk}$ denotes the rate of transitions (per cm$^3$ and s)
taking place from level $l$ to level $k$ and
$P_{lk} = R_{lk}+C_{lk}$.  This equation represents the
conservation equation for the particle number density $n_i$; however,
it also takes into account the generation and destruction of $n_i$
caused by transitions from and to other levels\footnote{Equation~(\ref{e54})
is in line with the earlier work by \cite{ramu03}.  In \cite{car02} based
on a 6-level atom it was shown that hydrogen ionization goes with
collisional excitation to $n=2$, then with radiative ionization
to the continuum, a pattern broadly consistent with the
present study.  On the other hand, \cite{car02} found that recombination
transitions occur by going from the continuum level to excited levels
with $n > 2$, which is not possible in the 3-level atom selected.}.
The LHS of
Eq.~(\ref{e54}) would be zero if the particles of species $i$ were
conserved; in this case, Eq.~(\ref{e54}) would represent the
standard continuity equation for particles of species $i$.

If time-dependent (i.e., noninstantaneous) ionization
processes are considered, the change in the population
density for the continuum level is given as
\begin{eqnarray}
\frac{dn_c}{dt} \ = \ \sum_{i=1}^{N-1} n_i ( R_{ic} + C_{ic} ) - n_c \sum_{i=1}^{N-1} ( R_{ci} + C_{ci} ) - n_{c}W
\label{eb10}
\end{eqnarray}
\citep{ramu03}; the latter also requires the time-dependent solution of the
advection term given as 
\begin{equation}
W \ = \ \rho\left( \frac{\partial T}{\partial p}\right)_S\left.\frac{dS}{dt}\right|_{\rm rad}
    - \frac{1}{\rho c_{\rm S}^{2}}\frac{dp}{dt} \ ,
\label{eb7}
\end{equation}
where $c_{\rm S}$ denotes the adiabatic sound speed given as
\begin{equation}
c_{\rm S} = \left[\left({\partial p\over \partial \rho}\right)_S\right]^{1/2} =
\left[
 \left({\partial \rho \over \partial p}\right)_T
+\left({\partial \rho \over \partial T}\right)_p
 \left({\partial T \over \partial p}\right)_S
 \right]^{-1/2}
\ .
\label{ea19}
\end{equation}

The number density of the hydrogen continuum population is closely related to the
electron density $n_e$, which is of critical importance for the manifestation of the
chromospheric radiative emission function \citep*[e.g.,][]{ver81,avr85}.  Other equations
concern the conservation of the total number of hydrogen particles $n_{\rm tot}$, given as
\begin{equation}
\frac{1+Z_{\rm el}}{X_{\rm el}} \sum_{i=1}^{N-1} n_i \ + \ \left( 1 + \frac{1+Z_{\rm el}}{X_{\rm el}} \right) \ n_c \ = \ \frac{p}{kT} \ = \ n_{\rm tot}
\ ,
\label{eb11}
\end{equation}
and the conservation of electrons,
given as
\begin{equation}
n_e \ = \ \left( 1 + \frac{Z_{\rm el}}{X_{\rm el}} \right) \ n_c \ + \ \frac{Z_{\rm el}}{X_{\rm el}} \sum_{i=1}^{N-1} n_i
\ ,
\label{eb12}
\end{equation}
where $X_{\rm el}$ and $Z_{\rm el}$ denote the element abundances of hydrogen and the metals,
respectively.  Note that $p$ is the gas pressure, $T$ is the temperature, and
$k$ is Boltzmann's constant.

\subsection{Comments on the (magneto-)hydrodynamic computer code}

The treatment of acoustic and longitudinal flux tube waves concerning
solar and stellar chromospheric models is pursued based on the
code by \cite{ramu03} and subsequent augmentations by \cite{ramc05b}.
This code is suitable for the simulation of one-dimensional
wave propagation and dissipation and, moreover, also allows
the treatment of time-dependent (i.e., non-instantaneous)
ionization processes for hydrogen and other elements by obtaining
solutions of the time-dependent statistical rate equations (see also
\citealt{car92}, \citeyear{car95}).  Additional features of the code
encompass the evaluation of the radiative losses (and gains)
by also taking into account departures from local thermodynamic
equilibrium (NLTE).  Shocks are treated as discontinuities based
on adequate solutions of the Rankine-Hugoniot relations; see, e.g.,
previous work by \cite{nie93} for a quasi-analytic method for the
computation of the thermodynamic relationships across shocks in
the presence of radiative and ionization processes.

The method of \cite{ramu03} is suitable for the treatment of both
monochromatic and spectral waves.  In the framework of our models,
time-dependent ionization is considered with respect to hydrogen
as well as magnesium and calcium.  Regarding the chromospheric
emission losses, we provide a detailed treatment of the Ca~II~K
and Mg~II~{\it k} lines; the total chromospheric emission losses
are obtained via appropriate scaling; see, e.g., \cite{cun99}
for previous applications of this approach.  The scaling factors
have been determined by inspecting representative solutions of
the wave models and by applying multi-level radiative transfer
computations with MULTI (see \citealt{ramf05}).  Radiative transfer
in the continuum, notable H$^-$, is pursued following the method
of \citet*{schm85}.  Another important aspect of our calculations is
the implementation of boundary conditions.  At the top boundary,
a transmitting boundary condition is used, whereas at the bottom
boundary, the injection of the magnetic or acoustic wave energy flux
is simulated via a piston-type boundary condition; the latter
is suitable for monochromatic waves as well as acoustic and
LTW frequency spectra; see \cite{ramu03} and references therein
for details.

\begin{figure}
\centering
\begin{tabular}{c}
\includegraphics[width=0.95\linewidth]{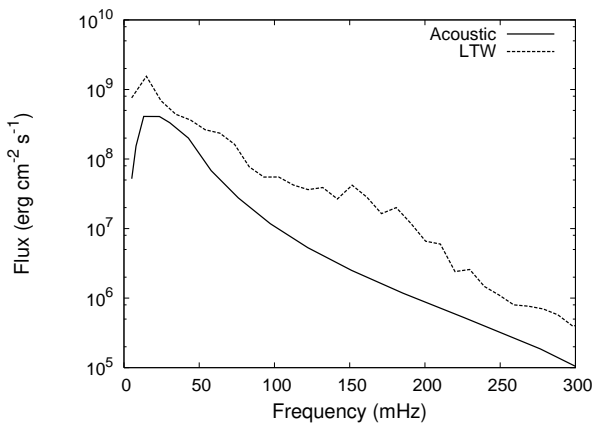}
\end{tabular}
\caption{Depiction of the generated flux for acoustic waves (solid)
and longitudinal flux tube waves (dashed) as function of frequency
for a mixing length of $\alpha_{\rm ML} = 1.8$.  For the magnetic waves
we consider the model with a bottom magnetic field strength of 1700 Gauss.
\label{fig1}}
\end{figure}

\begin{table*}
\caption{Acoustic and LTW energy generation.}
\begin{tabular}{lcccc}
\hline
$\alpha_{\rm ML}$ & \multicolumn{2}{c}{Acoustic} & \multicolumn{2}{c}{LTW} \\
\hline
... & $u_t$ & $F_{\rm M}$ & $u_t$ & $F_{\rm M}$ \\
... &  (cm s$^{-1}$) & (ergs cm$^{-2}$ s$^{-1}$) & (cm s$^{-1}$) & (ergs cm$^{-2}$ s$^{-1}$) \\
\hline
 1.5  &  $5.50 \times 10^4$ & $5.48 \times 10^7$ & $1.22 \times 10^5$ & $1.95 \times 10^8$  \\
 1.65 &  $6.81 \times 10^4$ & $7.85 \times 10^7$ & $1.25 \times 10^5$ & $2.48 \times 10^8$  \\
 1.8  &  $8.63 \times 10^4$ & $1.09 \times 10^8$ & $1.31 \times 10^5$ & $2.80 \times 10^8$  \\
 2.0  &  $1.09 \times 10^5$ & $1.54 \times 10^8$ & $1.35 \times 10^5$ & $3.02 \times 10^8$  \\
\hline
\end{tabular}
\end{table*}

\subsection{Wave energy fluxes and wave energy spectra}

The wave energy fluxes and wave energy spectra are calculated
following the approach by \cite{mus94} and \citet*{ulm96};
see also \cite{faw11} for updated results pertaining to a
grid of models for a set of main-sequence stars
including the Sun.  These types of models incorporate a
detailed description of the spatial and temporal spectrum of
the turbulent flow to obtain adequate results for the
frequency integrated acoustic energy fluxes along with the wave frequency
spectra; furthermore, they utilize an extended Kolmogorov spectrum
with a modified Gaussian frequency factor.
Both the acoustic and magnetic wave energy fluxes depend on the
mixing-length parameter $\alpha_{\rm ML}$, for which $\alpha_{\rm ML} = 1.8$
is used \citep{stei09a,stei09b}.

According to the updated value of $\alpha_{\rm ML}$, the
initial wave energy flux for the acoustic model is given as
$F_{\rm M} = 1.09 \times 10^8$ erg cm$^{-2}$ s$^{-1}$ \citep{ulm96};
see Table~1.  For the wave energy flux of the longitudinal flux tube
wave we use $F_{\rm M} = 2.80 \times 10^8$ erg cm$^{-2}$ s$^{-1}$
\citep{faw11}; this value assumes a pressure ratio between
outside and inside of the tube of $\epsilon = 3$ (see Eq.~\ref{e25}).
It is based on the assumption of an inside magnetic field strength of
1700~G, noting that in our model the equipartition magnetic field
strength is given as $B_{\rm eq} = \sqrt{8\pi p_e} = 2082$~G with
$B/B_{\rm eq} \simeq 0.82$.  The adopted inside tube magnetic field
strength has been taken at an optical depth of $\tau_{5000} = 1.3$,
corresponding to a mass column density of 2.1 g~cm$^{-2}$.  It is
found that the longitudinal wave energy flux is increased relative
to the acoustic energy flux by a factor of 2 to 3.5, depending on
the mixing-length parameter $\alpha_{\rm ML}$, owing to decisive
differences in the turbulent velocity $v_t$, which is considerably
enhanced in the magnetic case \citep[e.g.,][]{ulm98}.

For previous elaborations on the appropriate value for $F_{\rm M}$
and $B_{\rm eq}$ see, e.g., \citet*{ulm01}.
Detailed studies of magnetic field strengths for solar flux tubes
have been given by, e.g., \cite{sten78} and \cite{sol93}, allowing
us to motivate our choice of $B_{\rm eq}$.
Information on the wave frequency spectra for longitudinal
flux tube waves was presented by, e.g., \cite{ulm98} and \cite{ulm01};
see Fig.~1 for the spectral wave energy distribution serving
as basis for our models\footnote{The longitudinal wave energy spectrum
as obtained through the study of nonlinear time-dependent responses of
theoretical solar flux tubes to external pressure fluctuations
\citep[e.g.,][]{ulm98} is found to be somewhat bumpy, which is
a consequence of the occurrence of large-amplitude perturbations
(i.e., spiky waves) that are still apparent after temporal averaging
has been applied.}.  These authors found that the spectral energy
distribution for both the acoustic and longitudinal frequency spectrum
has a maximum close to a wave period of 60~s.  As our study also includes
monochromatic waves, this value will be used in the following for
both the acoustic and longitudinal wave study.

There is also considerable previous work on the most appropriate value
for the initial wave energy flux of longitudinal tube waves; the
latter is also modestly affected by the solar photospheric opacities
owing to their influence on the construction of the solar tube models
(i.e., attainment of radiative equilibrium, position of optical depth
$\tau_{5000}$).  Previous work by \cite{ulm01} adopted the opacity
tables compiled by \cite{boh84} and \cite{ulm96}, whereas more recent
simulations \citep{faw10} use the opacity table given by R. L. Kurucz
and collaborators (see \citealt{cas04} for details).

This latter approach is adopted
in the present paper.  In principle, the opacity table by R. L. Kurucz
and collaborators yields noticeably lower initial wave energy fluxes,
which for $\alpha_{\rm ML}$ between 1.8 and 2.0 are reduced by typically 30\%
compared to the models based on the opacity table considered by \cite{ulm01}.
A relatively high initial wave energy flux based on $\alpha_{\rm ML} = 2.0$
was also adopted by \cite{ramc05a}, which is a further reason for revisiting the
propagation and dissipation of longitudinal tube waves in the Sun by considering
an advanced treatment of the hydrodynamic and thermodynamic features as well as
a realistic initial wave energy flux.  Future investigations of stellar
convection and photospheres may also be based on the massively parallel
recently developed code Bifrost \citep{gud11}, which is designed to simulate
stellar atmospheres from the convection zone to the corona.

\begin{figure}
\centering
\begin{tabular}{c}
\includegraphics[width=0.95\linewidth]{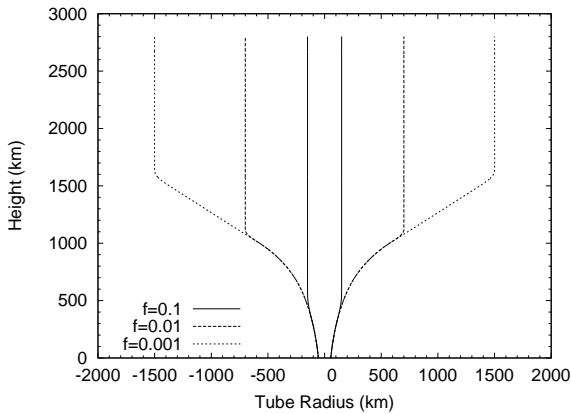}
\end{tabular}
\caption{Spread of magnetic flux tubes for the different magnetic
filling factors.
\label{fig2}}
\end{figure}

\begin{figure}
\centering
\begin{tabular}{c}
\includegraphics[width=0.95\linewidth]{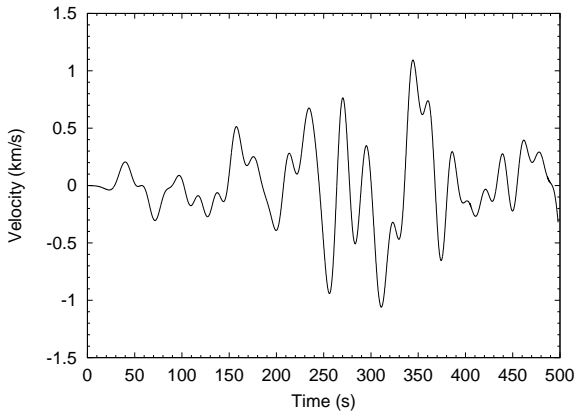}
\end{tabular}
\caption{Velocity at the bottom boundary of the flux tube models in response
to the longitudinal wave energy spectrum.
\label{fig3}}
\end{figure}

\subsection{Flux tube models}

For our models we consider three different flux tube geometries characterized
by different amounts of spreading as function of height; see, e.g., \cite{ram89}
and \cite{faw98} for previous work on the effects of tube spreading on the
deposition of non-radiative energy and the formation of chromospheric emission
with focus on the Sun.  The bottom radii of the tube models are chosen
as 55~km corresponding to half of the photospheric pressure scale-height;
note that the base pressure of the tubes is closely aligned to the VAL-C model
\citep{ver81}.  We also assume a magnetic field strength at the bottom of the 
tubes of $B_0 = 1700$~G informed by previous studies \citep{faw98}.  We
again consider wine-glass shaped tubes with top opening radii of 150~km,
700~km and 1500~km, respectively (see Table~2).
For these tubes, exponential spreading is assumed at relatively low heights,
followed by linear spreading.  Full spreading is attained at heights of
571~km, 1126~km and 1639~km, respectively; see Fig.~2 for a depiction of
the adopted tube models.

The tube with a top opening radius of 700~km has a top opening
area approximately twenty times larger than the tube with 150~km.
Moreover, the difference in the opening areas between the
wine-glass tubes with opening radii of 700 and 1500~km is
a factor of 4.6.  The flux tubes represent regions with magnetic
filling factors $f = r_{\rm B}^{2}/r_{\rm T}^{2}$ equal to 0.1\%, 0.01\%
and 0.001\%, respectively, with $r_{\rm B}$ and $r_{\rm T}$ denoting
the base radius and top radius of the tube, respectively. The
effect of flux tube spreading results in a progressively lower
inside magnetic field strength with a decreasing magnetic
filling factor $f$, entailing $B(r_{\rm T}) = 2.3$~G for
$f=0.001$ compared to 229~G for $f=0.1$ (see Table~2).

\begin{table}
\caption{Magnetic tube models.}
\begin{tabular}{lccc}
\hline
$f$  & $r_{\rm B}$ & $r_{\rm T}$ & $B(r_{\rm T})$ \\
(\%) & (km)        & (km)        & (G)            \\
\hline
 0.1   &  55  &  150  & 229   \\
 0.01  &  55  &  700  & 10.5  \\
 0.001 &  55  & 1500  &  2.3  \\
\hline
\end{tabular}
\end{table}


\begin{figure*}
\centering
\begin{tabular} {cc}
\includegraphics[width=0.45\linewidth]{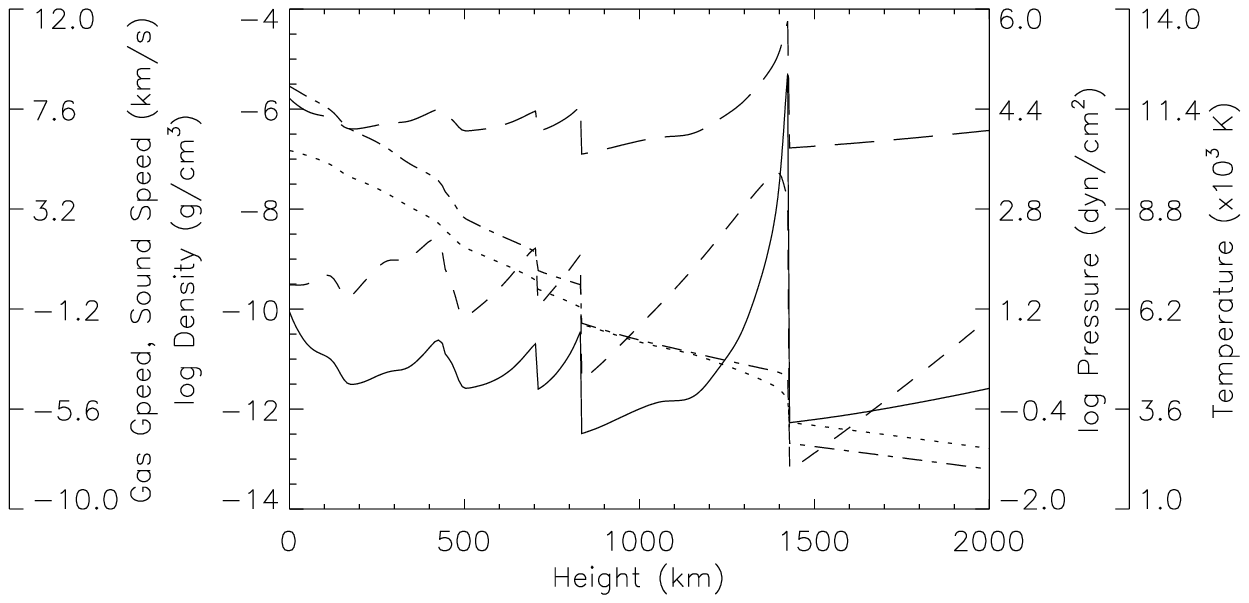} &
\includegraphics[width=0.45\linewidth]{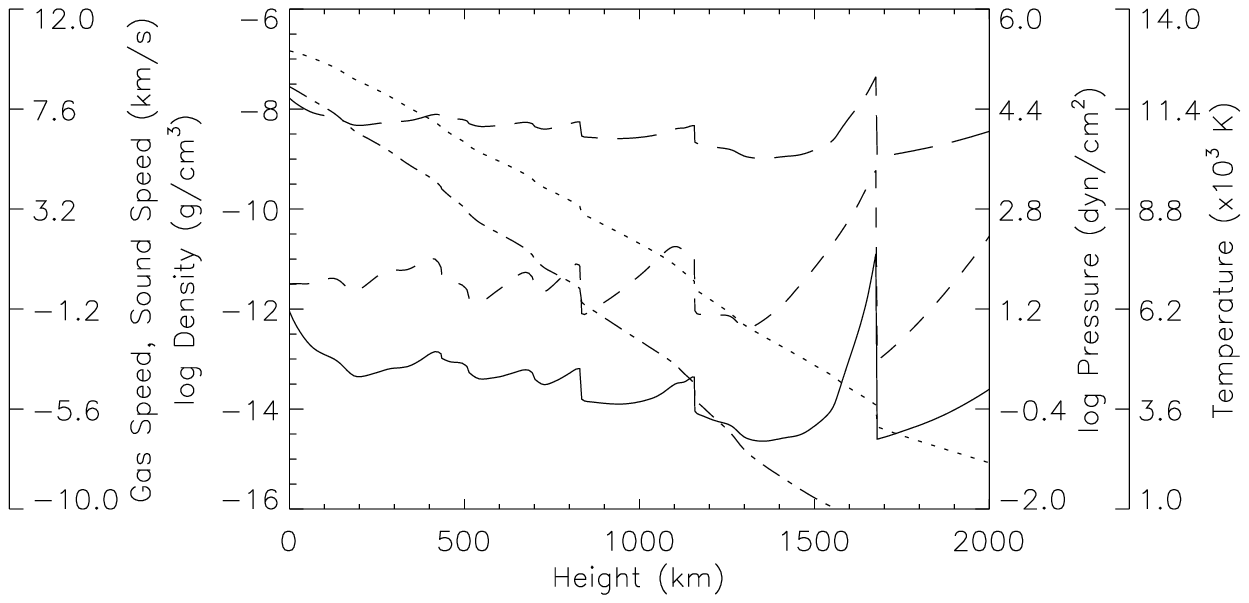} \\
\includegraphics[width=0.45\linewidth]{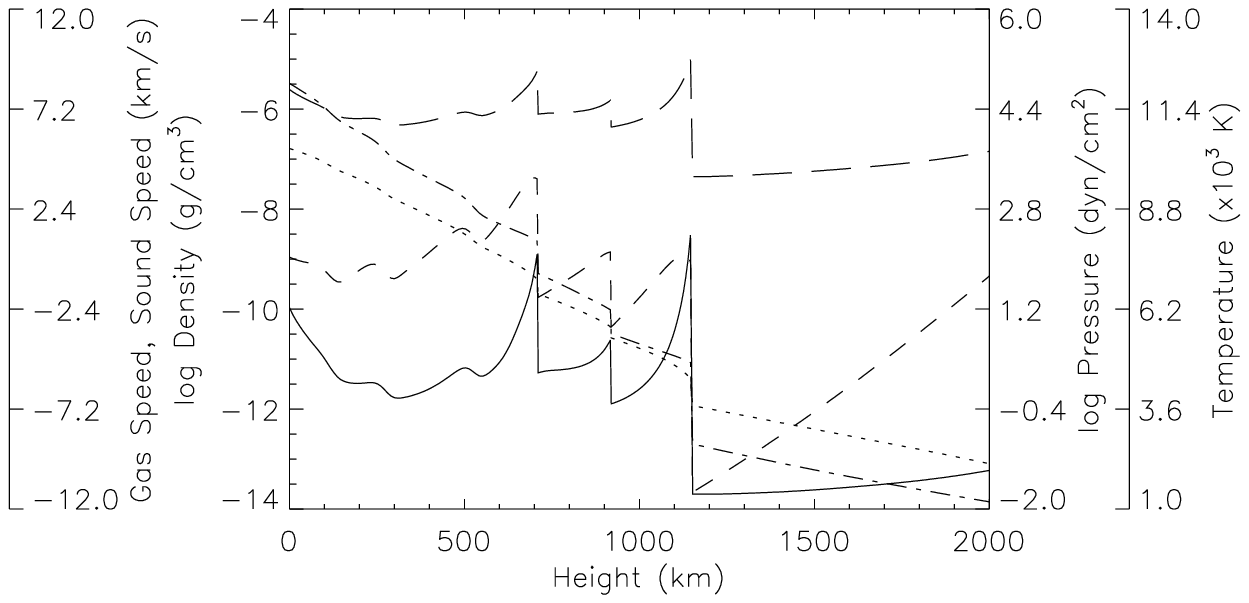} &
\includegraphics[width=0.45\linewidth]{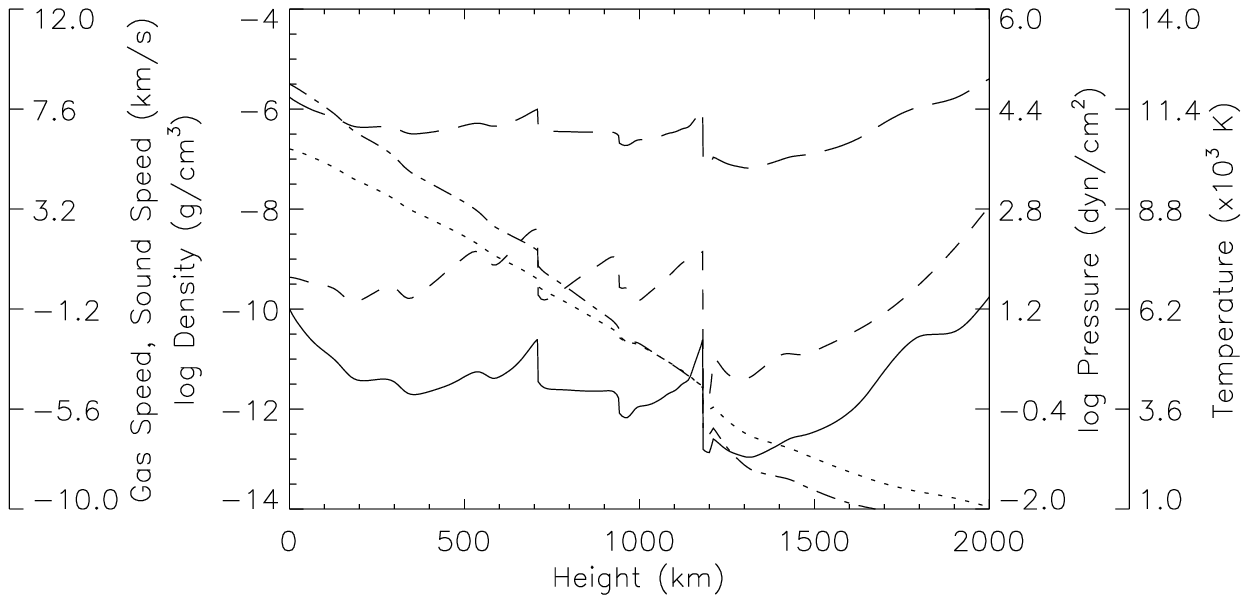} \\
\includegraphics[width=0.45\linewidth]{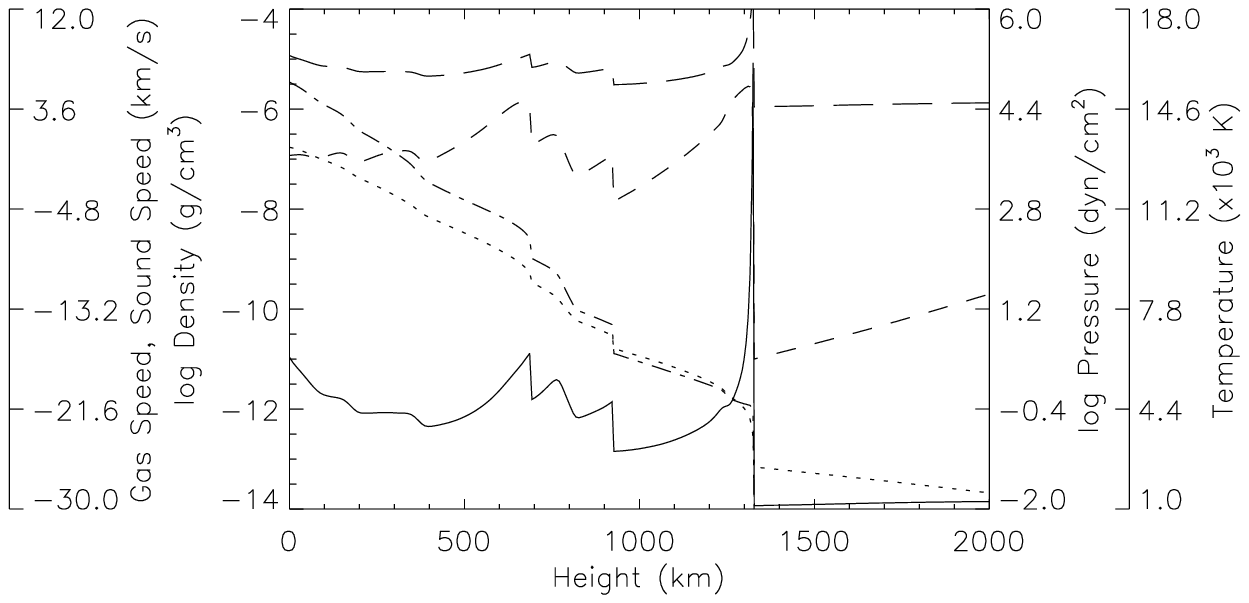} &
\includegraphics[width=0.45\linewidth]{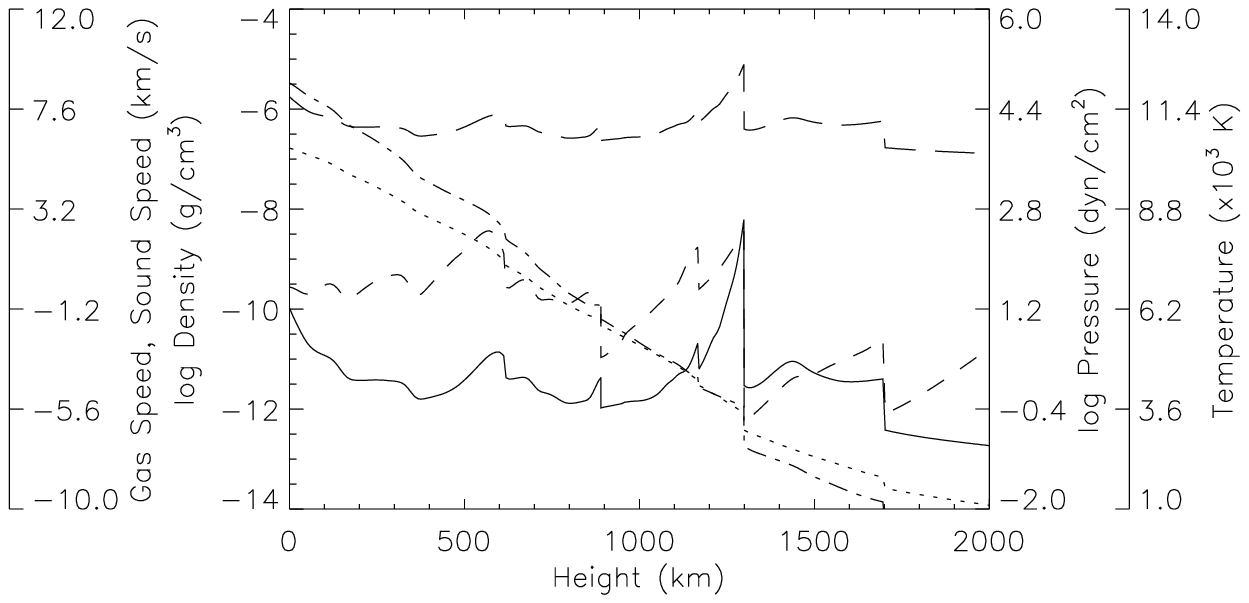} \\
\end{tabular}
\caption{
Two sequences of snapshots of longitudinal wave computations for magnetic flux
tubes with filling factors of $f = 0.1\%$ (left column) and $f = 0.001\%$
(right column) based on time-dependent ionization and with consideration
of frequency spectra.  These filling factors correspond to tube opening radii
of 150~km and 1500~km, respectively.
We depict the following variables: temperature (solid lines), gas speed
(dashed lines), density (short-dashed lines), sound speed (long-dashed lines),
and gas pressure (dashed-dotted lines).  The snapshots are taken at times
of 400, 800, and 1400~s (top to bottom).
}
\label{fig4}
\end{figure*}

\begin{figure*}
\centering
\begin{tabular} {cc}
\includegraphics[width=0.40\linewidth]{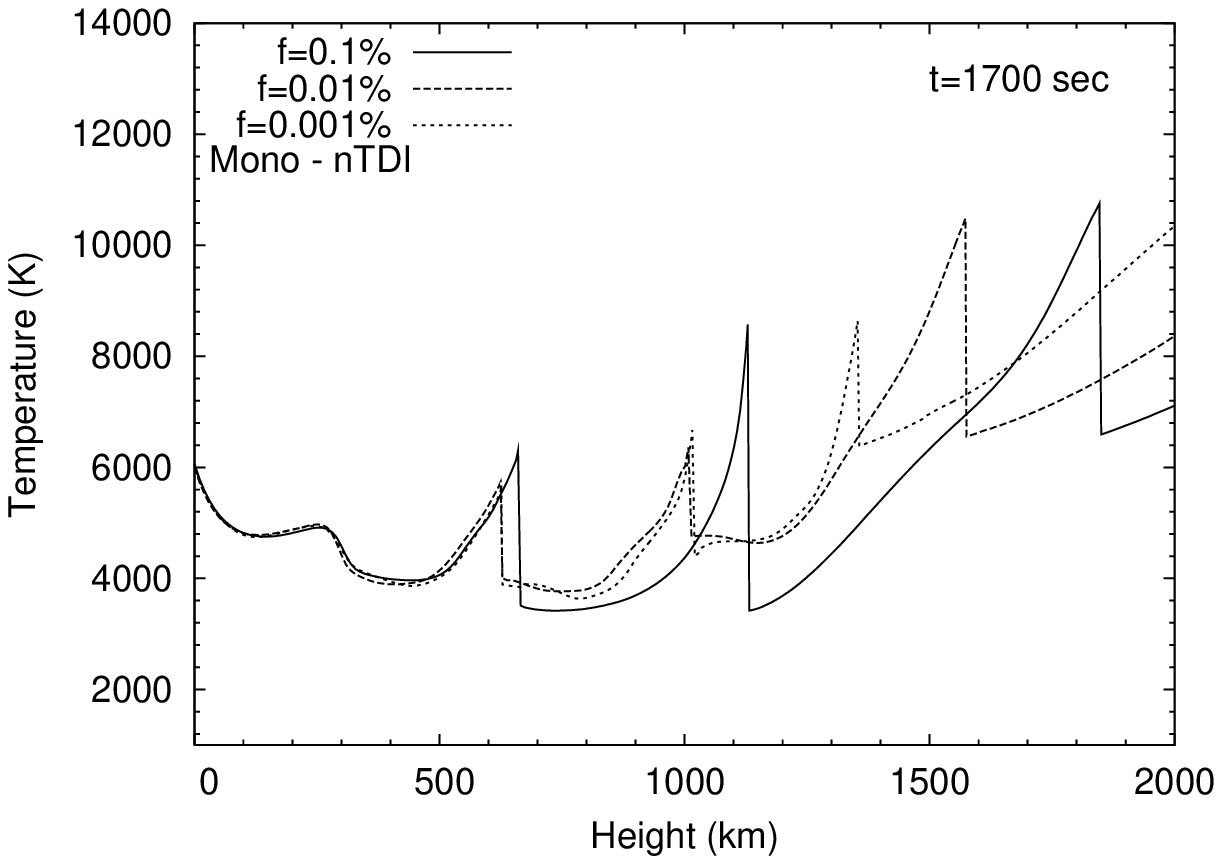} &
\includegraphics[width=0.40\linewidth]{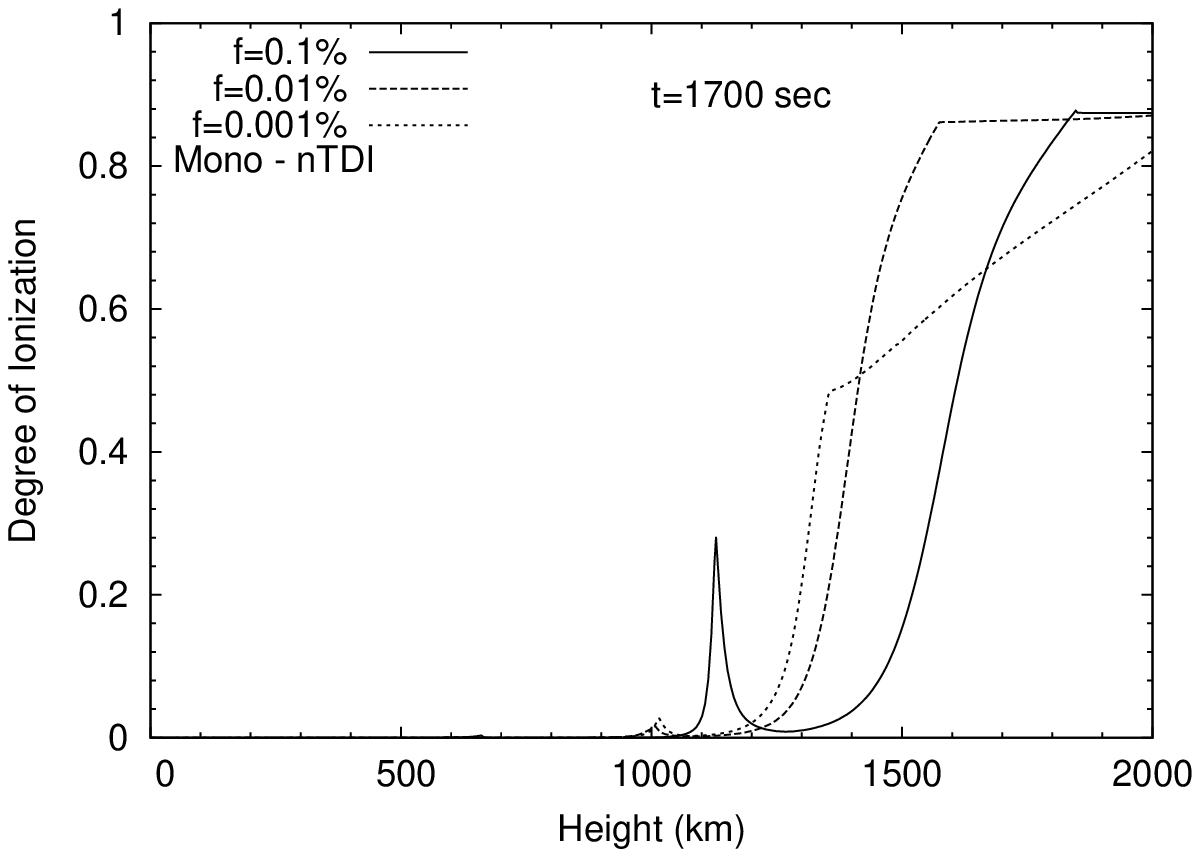} \\
\includegraphics[width=0.40\linewidth]{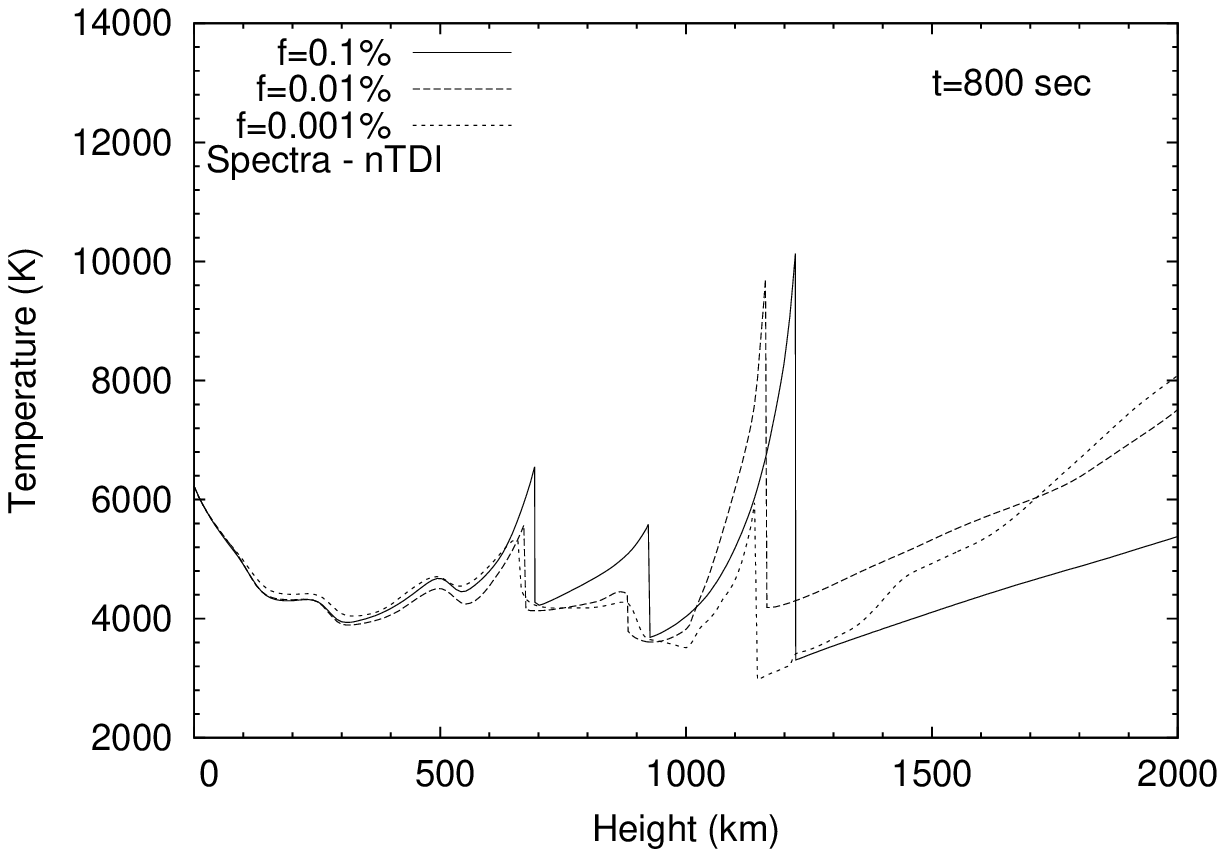} &
\includegraphics[width=0.40\linewidth]{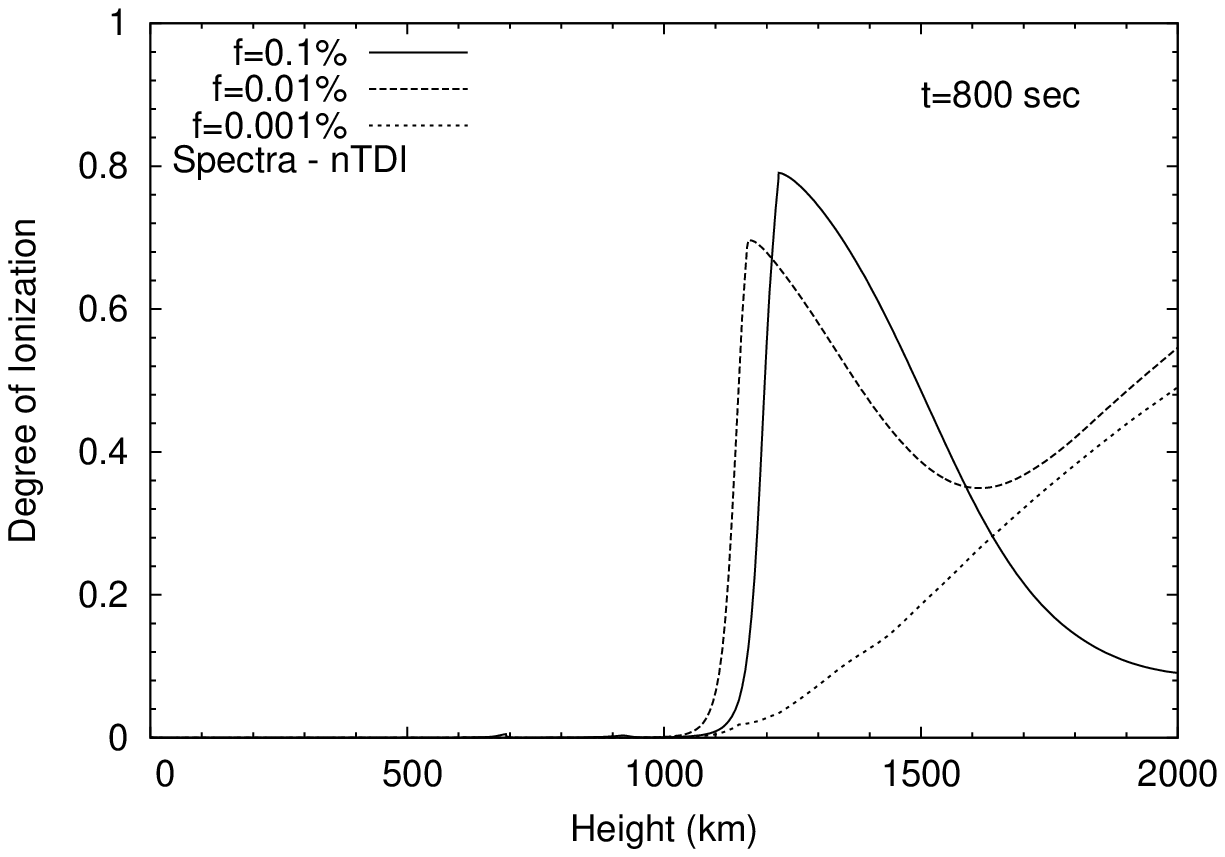} \\
\includegraphics[width=0.40\linewidth]{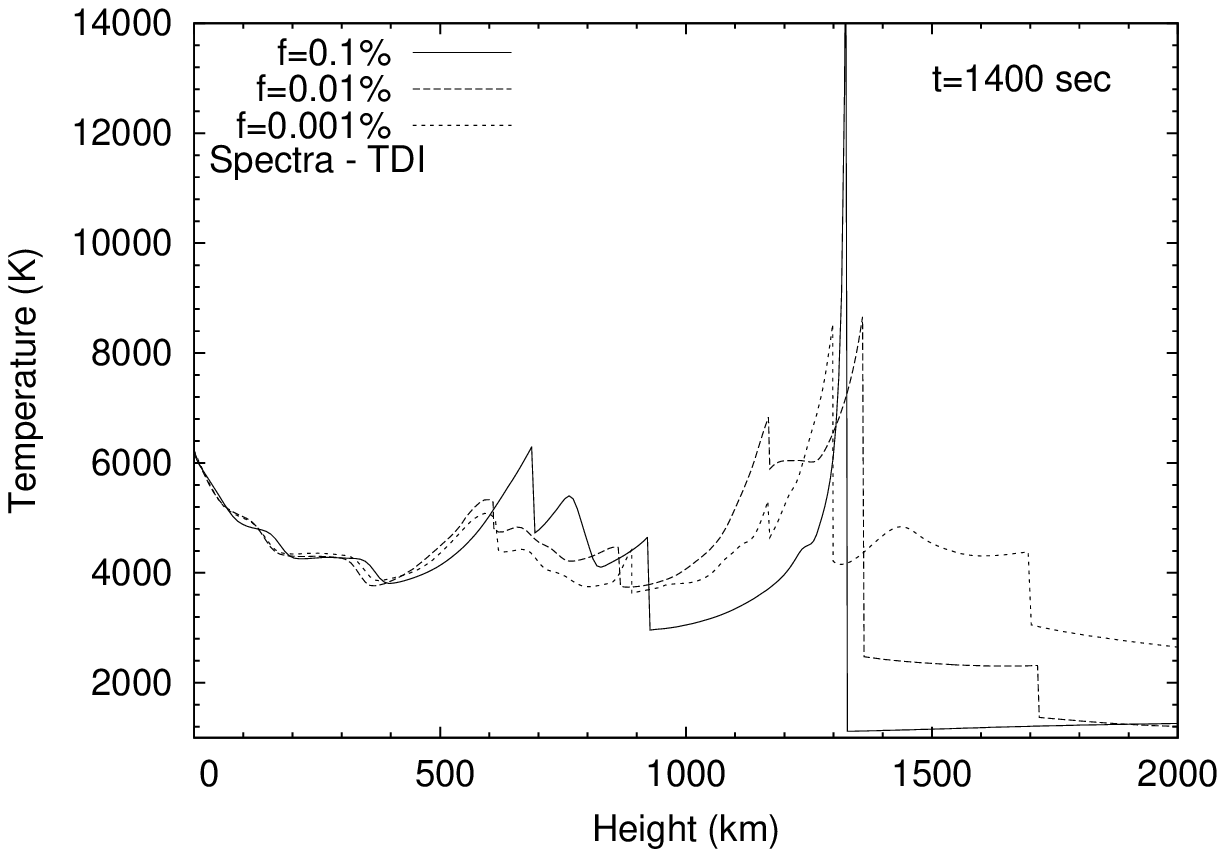} &
\includegraphics[width=0.40\linewidth]{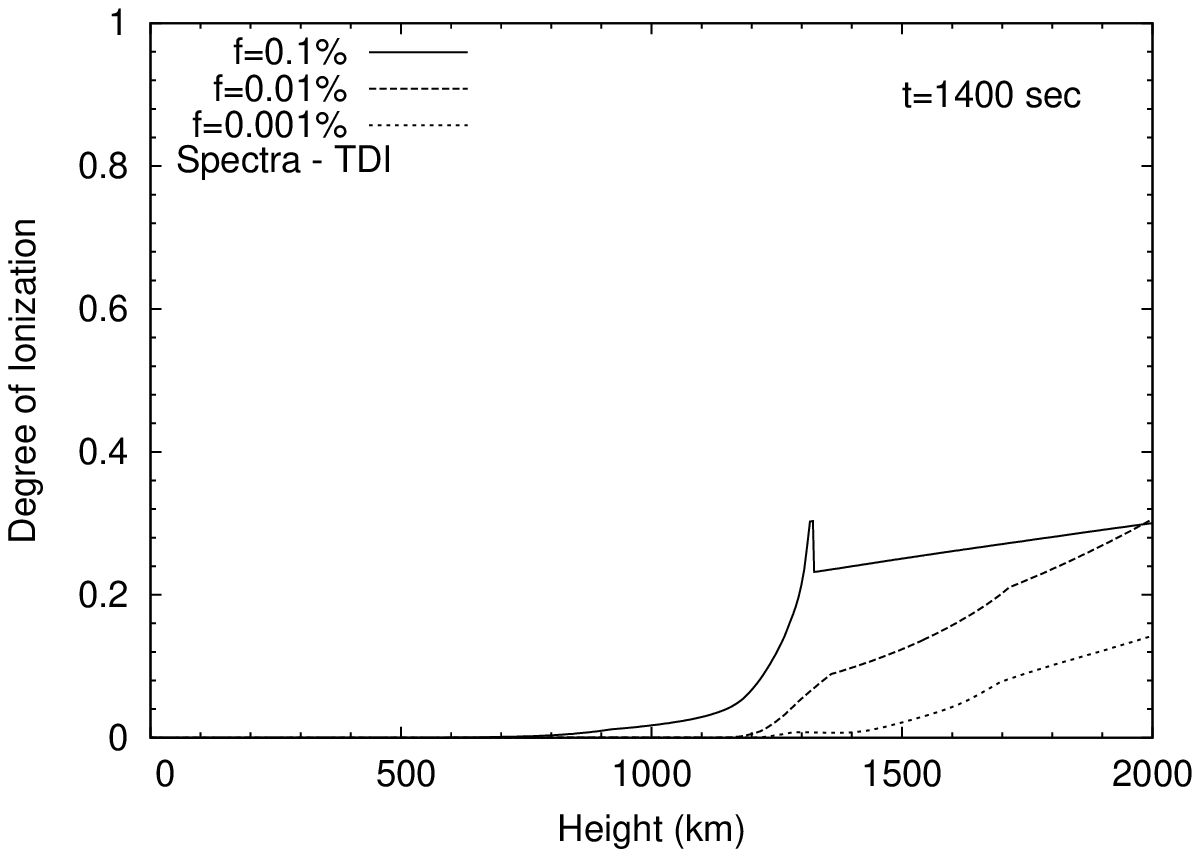} \\
\end{tabular}
\caption{
Behaviour of the temperature (left column) and hydrogen ionization
degree (right column) of propagating longitudinal flux
tube waves concerning tube models with different spreadings,
which correspond to filling factors of $f = 0.1\%$, 0.01\%,
and 0.001\%.  We depict monochromatic wave models (top) and
models with frequency spectra (middle) without time-dependent
ionization as well as models with frequency spectra including
time-dependent ionization (bottom).  Note the vast differences
in the temperature amplitudes.  The very large temperature spike
in the model with $f$ = 0.1\% (bottom) reaches 16,170~K.
}
\label{fig5}
\end{figure*}

\section{Results and Discussion}

\subsection{Detailed time-dependent model simulations}

By adopting the wave energy flux and wave frequency spectrum
for $\alpha_{\rm ML} = 1.8$, we pursued detailed simulations for
longitudinal tube waves for three different flux tubes
(see Sect. 2.4); in the following, we show calculations pertaining
to the magnetic filling factors of $f = 0.1\%$ and 0.001\%.
The stochstic velocity function at the bottom of the tubes
owing to the application of the longitudinal wave
frequency spectrum is depicted in Fig.~3.
Time-dependent ionization has been considered for
hydrogen as well as for Ca~II and Mg~II.  To demonstrate the
structure of our dynamic models in conjunction with the impact
of different tube spreadings, we depict snapshots at identical
elapsed times, which are: 400, 800, and 1400~s (see Fig.~4).

The general behaviour of the depicted waves in these models
is characterized by temporal increases of the wave amplitudes
resulting in the formation of shocks, typically occurring near 700
and 800 km for $f = 0.1\%$ and $0.001\%$, respectively.
The tube geometry determines the general position of the chromospheric
temperature rise.  It is found that the larger the tube spreading,
the greater the height of the temperature rise.  In models of
negligible tube spreading, the shock amplitudes tend to
stay constant or increase as function of height.  However, this
behaviour is counteracted by the impact of tube spreading: the
larger the spreading of the tube, the greater the area over which
the wave energy is distributed (i.e., geometrical dilution),
resulting in smaller wave amplitudes (see also Sect.~3.3).

Both models are shaped by strong interaction of shocks, which leads
to increased shock strengths $M_s$ as function of height. The shock
strength $M_s$ is given as
\begin{equation}
M_s \ = \ \frac{U_{\rm sh} - u_1}{c_{{\rm T}1}}
\ ,
\end{equation}
where $U_{\rm sh}$ denotes the shock speed, and $u_1$ and $c_{{\rm T}1}$
denote the gas speed and tube speed in front of the shock, respectively.

This behaviour is most pronounced for relatively narrow tubes; for
example, we find that the main shock at $t=1400$~s in the model of
$f = 0.1\%$ has a strength of 6.62 compared to 1.02 in the model of
$f = 0.001\%$.  This type of behaviour is also mirrored by the
jump in temperature across the shocks.  The corresponding
post-shock temperatures are typically considerably higher at
relatively large atmospheric heights and in models of relatively
narrow tubes, noting that narrow tubes imply higher shock strengths
anyhow and, moreover, entail more intense shock merging.
For $f = 0.1\%$ at an elapsed time of 1400~s
(see Fig.~4), the attained post-shock temperatures range between
4650 and 16,170~K.  In models of smaller magnetic filling factors,
the post-shock temperatures are typically considerably lower;
additionally, the variations in post-shock temperatures for
given models simulations and time steps is also reduced.
Furthermore, in models of time-dependent ionization, the post-shock
temperatures are also considerably higher than in equivalent models
without time-dependent ionization, a behaviour consistent with previous
acoustic models \citep[e.g.,][]{car92,car95,ramu03}; see also Sect.~3.2.

Additionally, the occurrence of stronger shocks in models
of $f = 0.1\%$ compared to $f = 0.001\%$ results in higher
levels of momentum transfer, which typically lead to smaller
densities in the top regions of those tubes as well as characteristic
behaviours for the time-averaged temperatures (see Sect.~3.3).
Although relatively large gas speeds may be attained by this process,
the outflow speeds typically remain subsonic.  Additionally,
due to energy dissipation and radiative energy losses, the wave energy
fluxes considerably decreases with atmospheric height (see Table 3
and 4).  However, comparing tubes of $f = 0.1\%$ and 0.001\%,
the decrease in the wave energy flux with height regarding the
tube with $f = 0.001\%$ is significantly less than expected from
the degree of geometrical dilution.  This phenomenon is caused by
the relatively low density in the tube with $f = 0.1\%$ owing
to the action of strong shocks (see Sect. 3.3); for previous
results on adiabatic LTW simulations see, e.g., \cite{faw98}.
At a height of 2000~km, the wave energy fluxes are given as
$5.10 \times 10^3$ and $1.93 \times 10^3$ ergs~cm$^{-2}$~s$^{-1}$
for $f = 0.1\%$ and 0.001\%, respectively (see Table 4).  For the
intermediate models of $f = 0.01\%$ (not shown), the wave energy
flux at 2000~km is given as $4.17 \times 10^3$ ergs~cm$^{-2}$~s$^{-1}$.
At low chromosphere heights, the heating and energy dissipation rates
are identified as almost independent of the magnetic filling factor
consistent with observational constraints \citep{bru93}.

\begin{table*}
\caption{LTW energy flux: monochromatic waves.$^a$}
\begin{tabular}{lcccc}
\hline
Height & \multicolumn{2}{c}{nTDI} & \multicolumn{2}{c}{TDI} \\
\hline
 ... & $f=0.1$ & $f=0.001$ & $f=0.1$ & $f=0.001$ \\
(km) &  ...    &  ...      &  ...    & ...       \\
\hline
 {\1p}600  &  1.43 $\times 10^7$  &  7.48 $\times 10^6$  &  1.08 $\times 10^7$  &  1.56 $\times 10^6$  \\
 {\1p}900  &  2.21 $\times 10^6$  &  4.15 $\times 10^5$  &  1.97 $\times 10^6$  &  1.01 $\times 10^5$  \\
     1200  &  2.62 $\times 10^5$  &  1.01 $\times 10^4$  &  2.56 $\times 10^5$  &  2.79 $\times 10^3$  \\
     2000  &  1.02 $\times 10^4$  &  3.37 $\times 10^3$  &  4.87 $\times 10^3$  &  2.67 $\times 10^2$  \\
\hline
\end{tabular}
\begin{enumerate}
\item[$^a$] The LTW energy fluxes are given in ergs cm$^{-2}$ s$^{-1}$.
\end{enumerate}
\end{table*}

\begin{table*}
\caption{LTW energy flux: spectral waves.$^a$}
\begin{tabular}{lcccc}
\hline
Height & \multicolumn{2}{c}{nTDI} & \multicolumn{2}{c}{TDI} \\
\hline
 ... & $f=0.1$ & $f=0.001$ & $f=0.1$ & $f=0.001$ \\
(km) &  ...    &  ...      &  ...    & ...       \\
\hline
 {\1p}600  &  1.09 $\times 10^7$  &  3.24 $\times 10^6$  &  1.53 $\times 10^7$  &  3.49 $\times 10^6$ \\
 {\1p}900  &  2.75 $\times 10^6$  &  4.90 $\times 10^5$  &  3.33 $\times 10^6$  &  5.49 $\times 10^5$ \\
     1200  &  4.30 $\times 10^5$  &  3.31 $\times 10^4$  &  5.07 $\times 10^5$  &  2.58 $\times 10^4$ \\
     2000  &  4.34 $\times 10^4$  &  5.41 $\times 10^3$  &  5.10 $\times 10^3$  &  1.93 $\times 10^3$ \\
\hline
\end{tabular}
\begin{enumerate}
\item[$^a$] The LTW energy fluxes are given in ergs cm$^{-2}$ s$^{-1}$.
\end{enumerate}
\end{table*}

\subsection{Effects of flux tube spreading and time-dependent ionization}

To obtain further insight into the effect of tube spreading on the
atmospheric shock strengths as well as into other features owing to
time-dependent ionization we computed a set of detailed models.
They include monochromatic wave models without time-dependent ionization
as well as models with frequency spectra with and without time-dependent
ionization.  For each set of models, we considered tube spreadings
corresponding to magnetic filling factors of $f = 0.1\%$, 0.01\%,
and 0.001\%, respectively, thus allowing us to gauge the impact of
tube spreading on the temperature structure and shock strengths.

The first set of models is based on monochromatic waves.
For all three tube spreadings we chose an elapsed time of 2500~s,
thus ensuring appropriate comparisons between the models.  Hence,
almost 30 shocks have been inserted into the atmosphere; therefore,
all three tube atmospheres have reached a dynamic steady-state, i.e.,
the interaction between the shocks (i.e., ongoing shock merging) has
subsided.  Comparions between models of different tube spreading
(see Fig.~5; top panel) indicate that the models with
narrow tube spreading (i.e., high magnetic filling factor) are
characterized by relatively high shock strengths, whereas models
with wide tube spreading (i.e., low magnetic filling factor)
are shaped by relatively small shock strengths.  Specifically, for
regions beyond 1200~km, the typical shock strength for $f = 0.1\%$
varies between $M_s = 1.85$ and 2.77, whereas for $f = 0.001\%$,
it is only $M_s = 1.60$.  For $f = 0.01\%$, intermediate shock strength
values are found, as expected.  In models of monochromatic waves
the shocks have attained limiting shock strength, which is found
to depend on the height-dependent behaviour of the tube spreading,
as previously pointed out through analytical means \citep{cun04}.
Previous simulations for solar flux tubes based on adiabatic LTW
waves without time-dependent ionization also show a similar kind
of behaviour \citep{faw98}.  Due to the omission of detailed
time-dependent ionization, it is found that the degree of hydrogen
ionization at 2000~km is about 95\% for $f = 0.1\%$ and $0.01\%$.
For $f = 0.001\%$, a lower value for the hydrogen ionization degree
is found; in this case, a smooth and steady increase of the hydrogen
ionization degree occurs between 1200 and 2000~km.

The second and third sets of models are based on LTW frequency spectra.
Figure~5 (middle and bottom panel) shows snapshots without and with the
consideration of time-dependent ionization, respectively.  For both types
of models, we again consider magnetic filling factors of $f = 0.1\%$,
0.01\%, and 0.001\%.  The snapshots displayed for these sets of models
are taken at 800 and 1400~s, respectively; furthermore, the elapsed
time of simulation is about 2000~s.  The main shocks in Fig.~5
(middle panel), attained through the process of shock merging, have
strengths of 3.19, 3.18, and 2.11, respectively.  The main shock for
$f = 0.1\%$ has a post-shock temperature of 10,070~K; the hydrogen
ionization degree in front of the shock is 78\%, and shortly behind the
shock it is close to zero.  By comparison, the post-shock temperatures
of the main shocks pertaining to $f = 0.01\%$ and 0.001\% are 9250~K
and 6470~K, respectively, which clearly indicates the impact of tube
spreading.  Clearly, the change of structure due to different magnetic
filling factors is a direct consequence of the dilution of the wave
energy flux (see Table 4), including associated magnetohydrodynamic,
thermodynamic and radiative phenomena, particularly those associated
with the formation of shocks of different strengths.  At heights of
1200~km, the wave energy flux is reduced by factors of
$1.5 \times 10^{-3}$, $2.9 \times 10^{-4}$, and $1.9 \times 10^{-5}$
in tube models with $f = 0.1\%$, 0.01\%, and 0.001\%, respectively.

Figure~5 (bottom panel) reflects a similar setting but now
both time-dependent ionization and LTW frequency spectra
are taken into account.  The main difference to the kind
of simulations previously discussed is an enhanced tendency
of building up very strong shocks due to shock merging,
particularly for the model with a
magnetic filling factor of 0.1\%.  In this case, a very large
temperature spike with a post-shock temperature of 16,170~K
is found to occur; the corresponding shock strength is given
as $M_s = 6.62$.  In contract, however, at relatively low
heights, a considerable number of small shocks is encountered.
Moreover, at heights between 600 and 1000~km, shock strengths
are typically between $M_s = 1.2$ and 1.5 in models of
$f = 0.1\%$.  The shock strengths are even lower in models
with 0.001\%; in the latter cases, the shock formation is
postponed to larger heights.  The merging of shocks leads
to the build-up of very strong shocks, a process that is
significantly amplified in models of time-dependent ionization,
owing to the fact that virtually no energy is used immediatelly
behind the shocks to ionize hydrogen (or any other species).
Therefore, an increased amount of energy is available to further
increase the post-shock temperatures.

\begin{figure}
\centering
\begin{tabular} {c}
\includegraphics[width=0.95\linewidth]{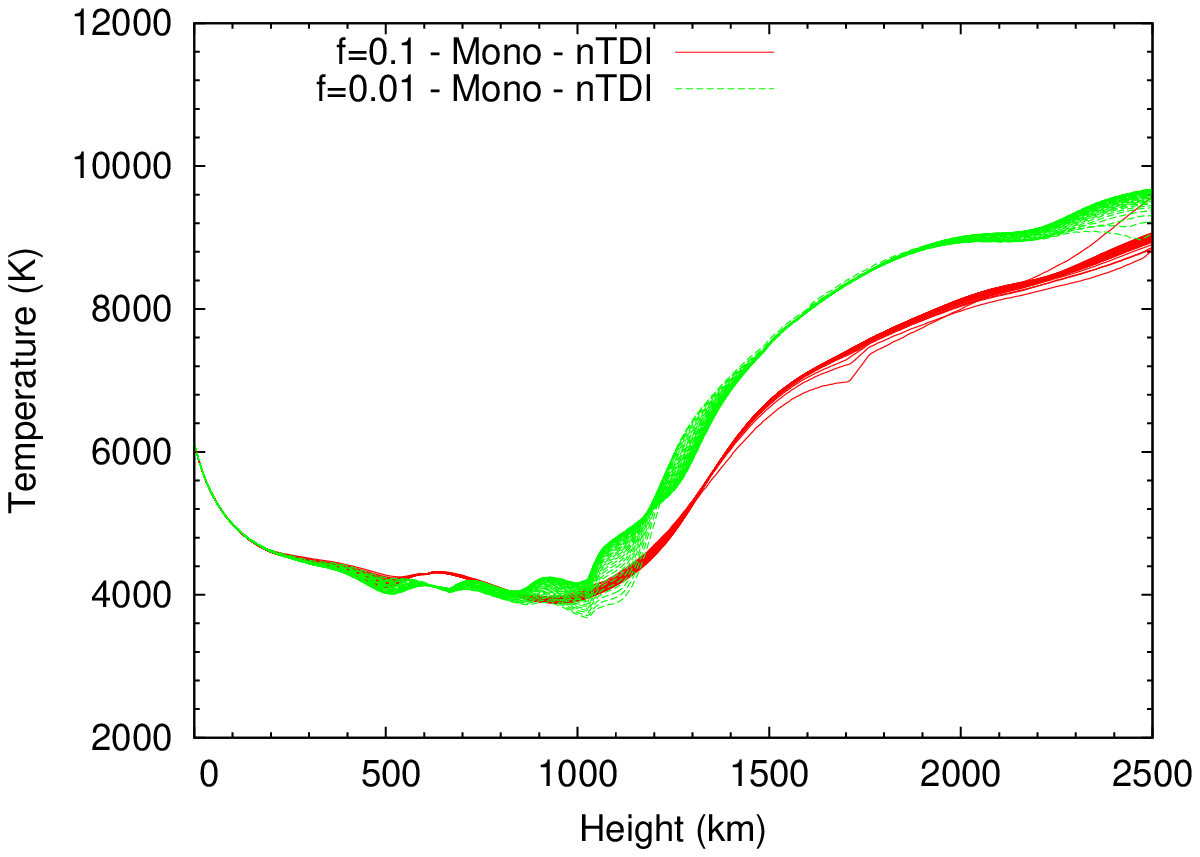} \\
\includegraphics[width=0.95\linewidth]{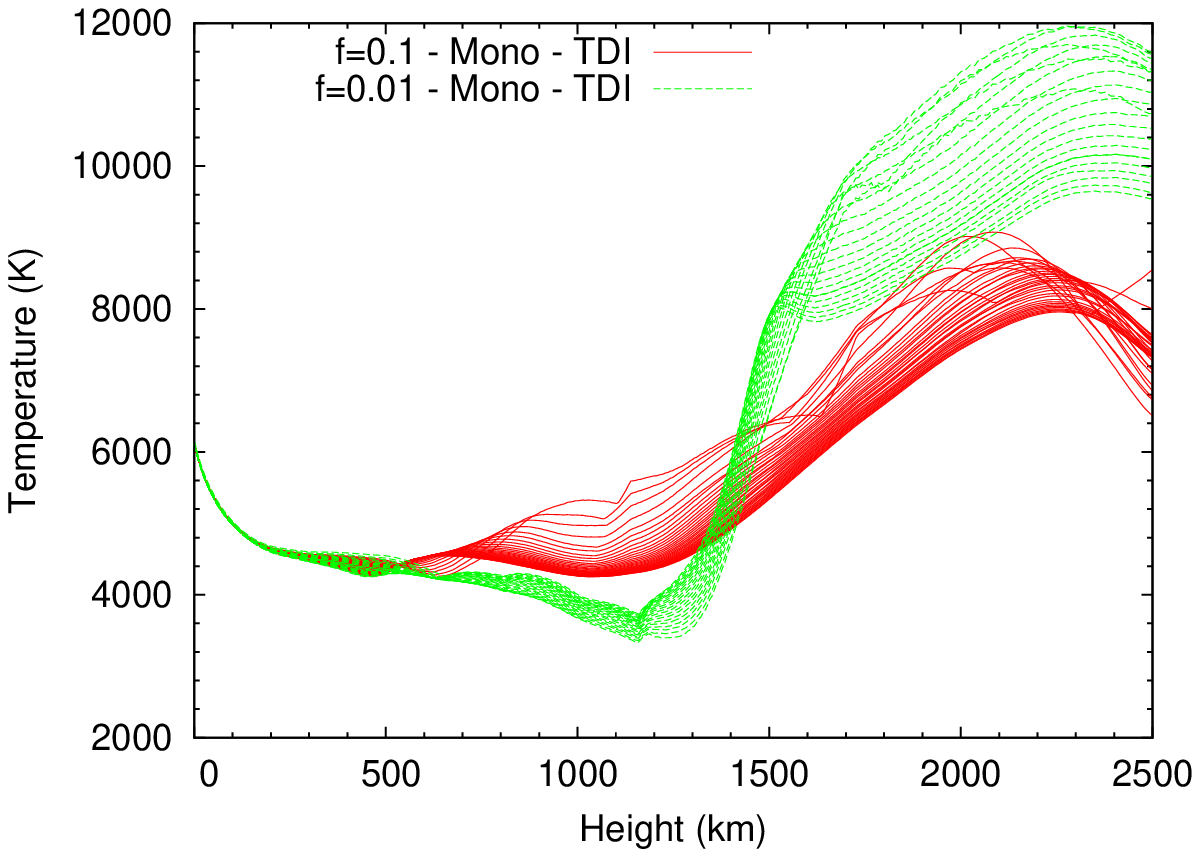} \\
\end{tabular}
\caption{
Build-up of mean (i.e., time-averaged) temperatures for
models with and without time-dependent
ionization (bottom and top figure, respectively).  Note
the influence of the different tube spreadings, denoted
as magnetic filling factors $f = 0.1\%$ and 0.01\%; see
Table~2.  The time of averaging for the various models
is close to 60~s.  For the simulations with time-dependent
ionization the averaging starts at 513~s ($f = 0.1\%$) and
821~s ($f = 0.01\%$) and for the simulations without
time-dependent ionization it starts at 607~s ($f = 0.1\%$)
and 612~s ($f = 0.01\%$).
}
\label{fig6}
\end{figure}

The dilution of the wave
energy flux is also highly significant in these types of models
depending on the degree of geometrical spreading.  At heights
of 1200~km, the initial wave energy flux is reduced by factors
of $1.8 \times 10^{-3}$, $3.9 \times 10^{-4}$, and $9.2 \times 10^{-5}$,
in models with $f = 0.1\%$, 0.01\%, and 0.001\%, respectively.
However, an assessment of the absolute amount of wave energy
flux indicates (see Table 3 and 4) that it is less diluted 
than expected from the degree of geometrical dilution, which
obviously is a factor of 100 between $f = 0.1\%$ and 0.001\%.
The reason for this type of behaviour is that models with
$f = 0.1\%$ have a lower overall density; furthermore, in
those models there is a higher loss of energy behind the
shocks due to enhanced radiative energy losses and the
initiation of episodic outflow.  For the same reason, the
reduction of the wave energy flux with height is also more
drastic in models with time-dependent ionization compared to
models without consideration of this process (TDI versus nTDI).

For the assortments of small shocks, it is noteworthy that different
shock strengths correspond to different shock speeds; therefore, the
shocks in models with LTW frequency spectra are poised to overtake
one another, thus invoking significant shock merging as well as the
build up of very strong shocks.  This result has already
been obtained in 1-D acoustic models \citep[e.g.,][]{cun87},
and it is evident in 1-D LTW models with and without the consideration
of time-dependent ionization, but it is drastically enhanced if
time-dependent ionization of hydrogen is considered.  The
occurrence of very strong shocks is a typical feature of the combined
influence of frequency spectra and time-dependent ionization inherent
in 1-D (magneto-) hydrodynamic models; see, e.g., \cite{car92} for
previous results on acoustic models.  Figure~5 (middle and bottom panels)
also depicts the behaviour of the hydrogen ionization degree in the
various models in front and behind the shocks.  Clearly, the hydrogen
ionization degrees are progressively lower in models of smaller values
of $f$; however, in models without time-dependent ionization of hydrogen
it drops to almost zero in the post-shock regions (see Fig.~5, middle
panel), whereas in models with time-dependent ionization of hydrogen it
is essentially maintained (see Fig.~5, bottom panel).

\begin{figure*}
\centering
\begin{tabular} {cc}
\includegraphics[width=0.40\linewidth]{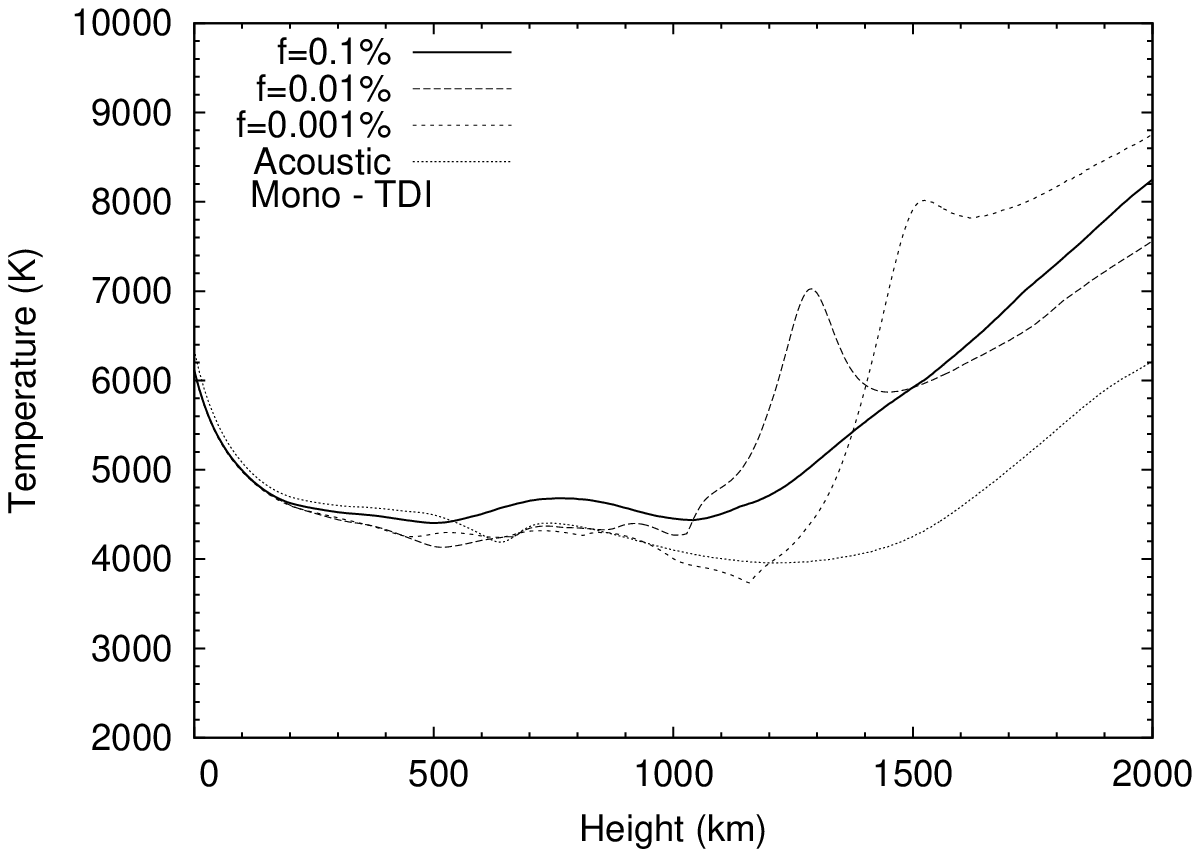} &
\includegraphics[width=0.40\linewidth]{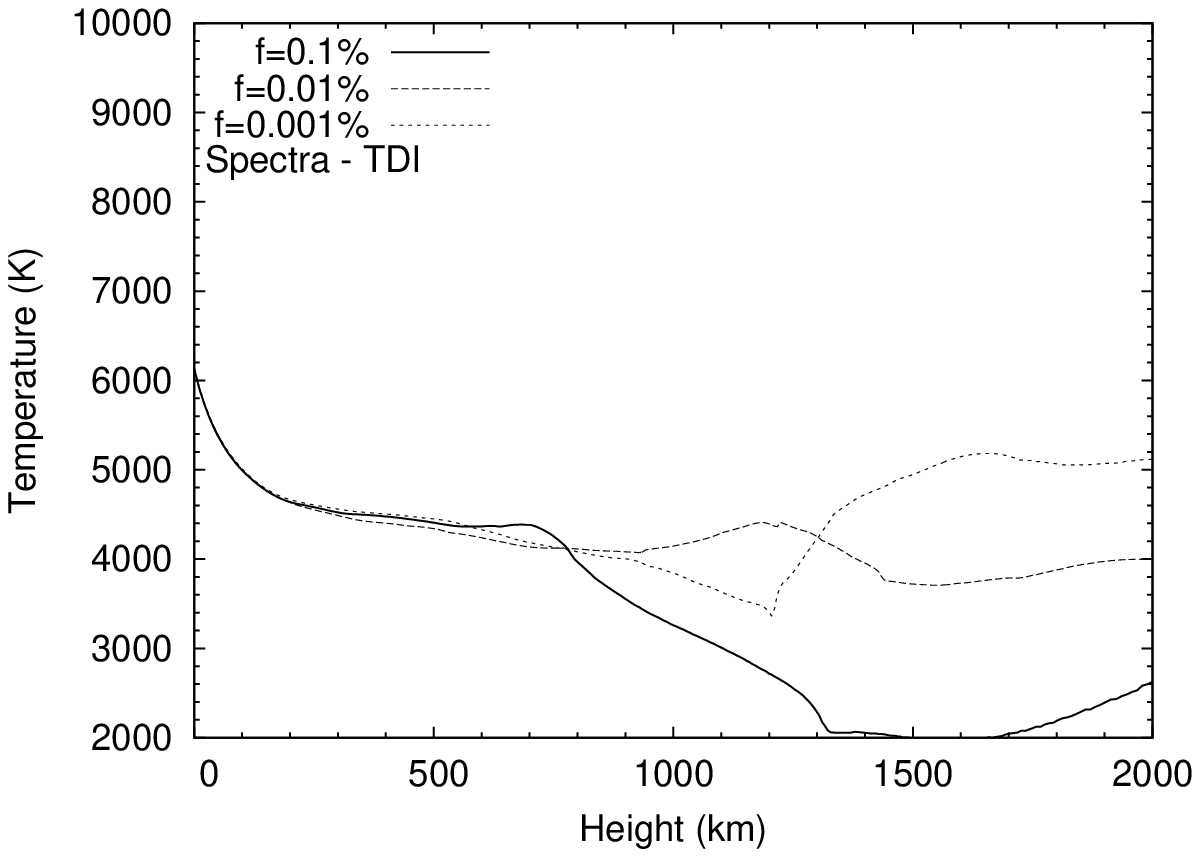} \\
\includegraphics[width=0.40\linewidth]{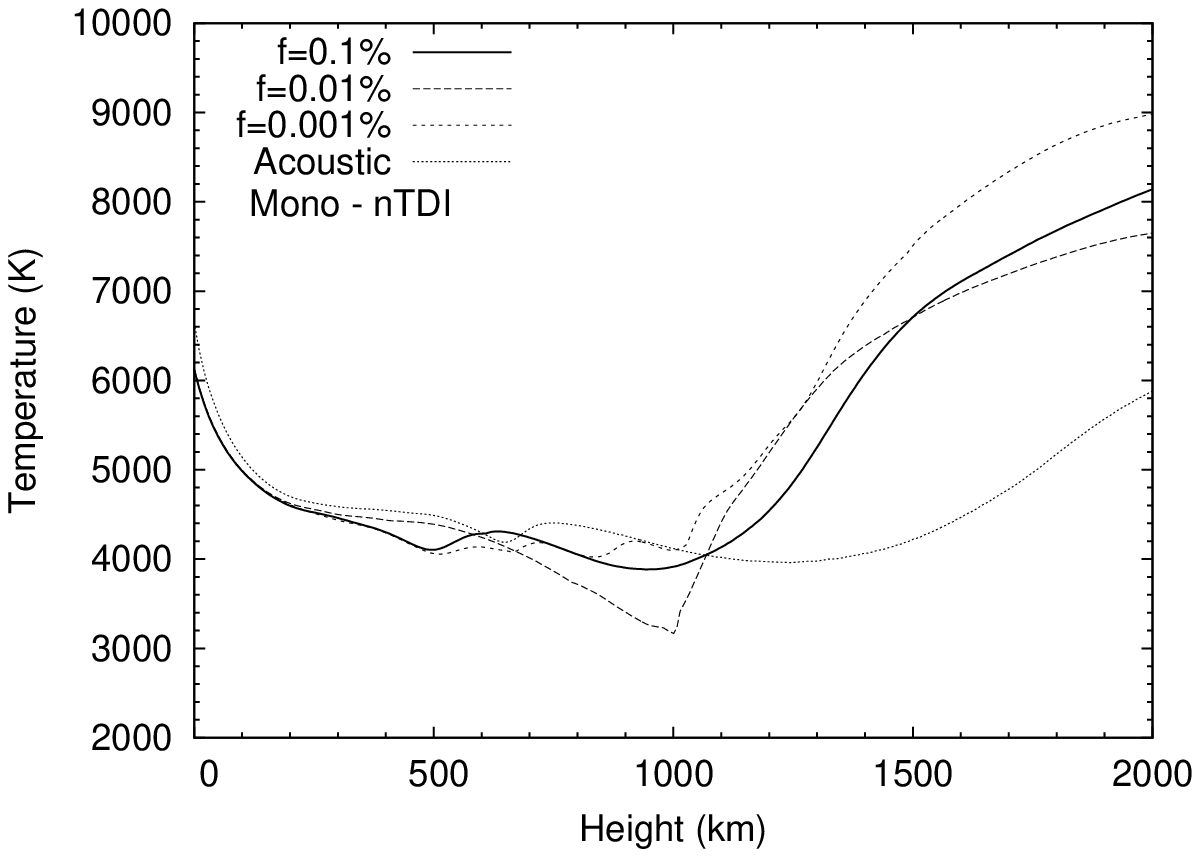} &
\includegraphics[width=0.40\linewidth]{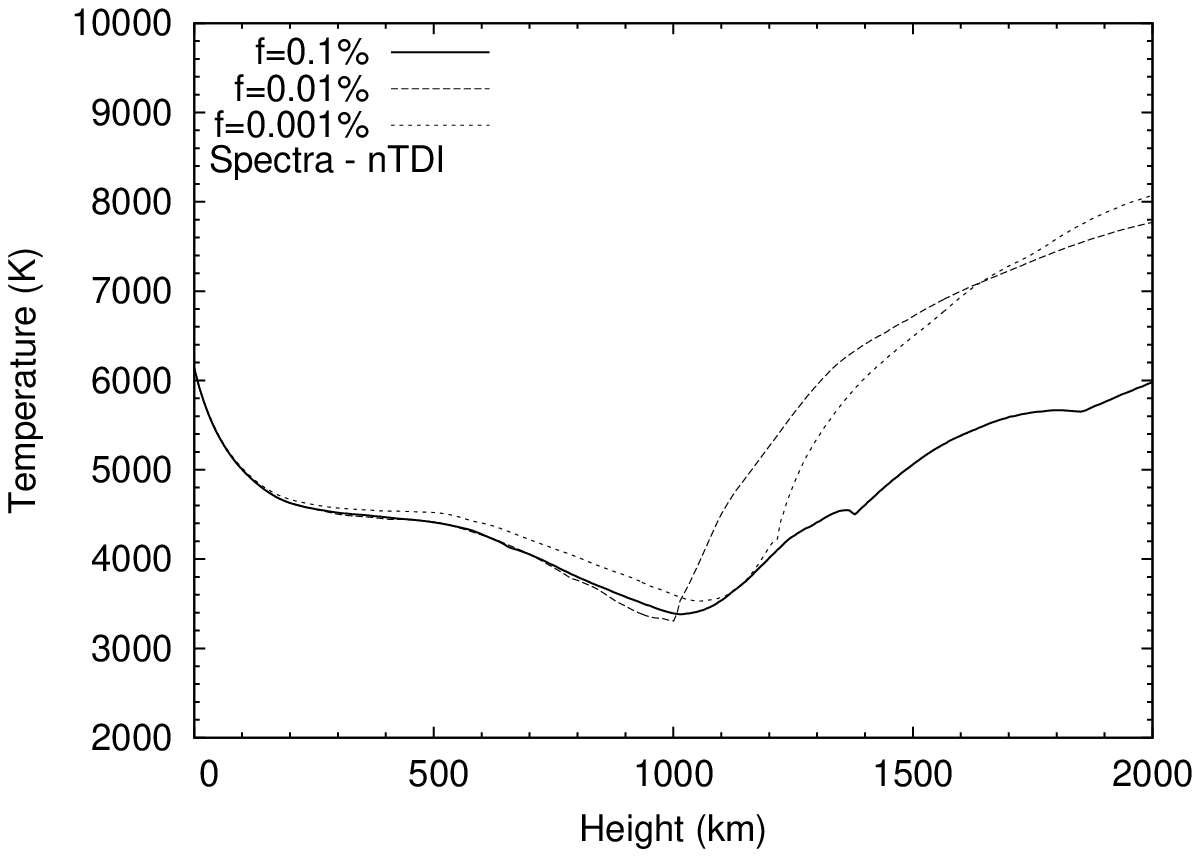} \\
\end{tabular}
\caption{
Comparison of time-averaged temperatures for longitudinal wave
computations pertaining to different types of models.
We show models based on monochromatic waves (left column) and
frequency spectra (right column).  The top and bottom row
depicts models with and without consideration of time-dependent
ionization, respectively.  Concerning the model series based
on monochromatic waves, we also show acoustic wave models for
comparison.  Regarding the magnetic flux tubes models, we
show simulations for magnetic filling factors of $f = 0.1\%$,
0.01\%, and 0.001\%.
}
\label{fig7}
\end{figure*}

\subsection{Behaviour of time-averaged temperatures and ionization degrees}

Additional insight into the overall dynamic structure of
flux tubes subjected to the propagation and dissipation of
longitudinal tube waves can be obtained by assessing the
behaviour of time-averaged temperatures.  Again, we considered both
monochromatic and spectral waves, while pursuing detailed
comparisons for models with and without time-dependent ionization.
Concerning monochromatic models, we also pursued computations
for acoustic waves aimed at representing atmospheric structure
exterior to the flux tubes; in this regard, plane-parallel
geometry was assumed.  In all types of models, the wave energy
generation was derived from a mixing length of $\alpha_{\rm ML}
= 1.8$ resulting in an initial wave energy flux of
$1.09 \times 10^8$ and $2.80 \times 10^8$ erg~cm$^{-2}$~s$^{-1}$
for the acoustic and magnetic models, respectively (see Table~1).
For the magnetic flux tube simulations, we again considered
tube geometry given by magnetic filling factors of $f = 0.1\%$,
0.01\%, and 0.001\%.  Examples for the steady build-up of
time-averaged temperatures owing to the implemented averaging
procedure are conveyed in Fig.~6.

Detailed information on the behaviour of the
time-averaged temperatures is given in Fig.~7, which depicts
a total of 14 models.  Let us first focus on the evaluation of
dynamic magnetic flux tube models based on monochromatic waves.
Typically, we awaited the insertion of 20 to 50 wave periods
into the atmosphere prior to the evaluation of mean (i.e.,
time-averaged) quantities to ensure that the switch-on behaviour
of the tube atmospheres has subsided and dynamic equilibria have
been reached.  The time-averaged temperatures of the various tube
models as well as those of the acoustically heated models
are relatively similar at heights below 800 km, although they
are found to be highest in the model with the smallest spreading
(i.e., $f = 0.1\%$), as expected.  This latter result can be understood
based on the dilution of the wave energy flux, which also affects
the height of shock formation.  The latter is given as 400, 550, and
850~km for the models with 0.1\%, 0.01\%, and 0.001\%, respectively.
For the acoustic model, the height of shock formation is given as
600~km.  Only minor differences are obtained between models with
and without time-dependent ionization, noting that due to the
relatively high densities, low temperatures and small shock
strengths (if any), the effects due to time-dependent ionization
are less pronounced.

In the upper parts of the tube atmospheres, the temperatures
in the model of the widest tube opening radius are expected to be
lowest due to the largest dilution of the wave energy flux.
The difference in dilution is expected to be a factor of 100
between $f = 0.1\%$ and 0.001\%; however, it is significantly
lower, i.e., a factor between 3 and 26 for models with or
without time-dependent ionization and with or without the
consideration of LTW frequency spectra (see Table~3 and 4).
For some portions of the tube atmospheres, it is even found that
``the larger the tube opening radius, the higher the mean
temperatures", a result that is highly counter-intuitive.
The reason for this behaviour is that the dilution of the wave
energy flux does not exactly follow geometrical scaling.
Other relevant processes involve the density structure within
the flux tubes as well as significant dynamical and radiative
processes, especially pertaining to strong shocks, including
significant radiative energy losses.   Comparing models with
and without time-dependent ionization reveals that time-dependent
ionization leads to lower average temperatures inside the flux
tubes compared to time-independent ionization.  However, the
average temperatures inside of flux tubes in models with or
without time-dependent ionization are nonetheless noticeably
higher than in the acoustically heated external atmosphere,
pointing to the significance of magnetic heating.

We also pursued corresponding sets of models based on LTW frequency
spectra instead of monochromatic waves.  In this type of models the
attained time-averaged temperatures were unrealistically low,
particularly in models of time-dependent ionization.  In this case,
the combined effects of time-dependent hydrogen ionization and of the
dynamical structure related to the propagation of spectral waves lead
to very strong shocks, which invoked amplified energy losses behind
the shocks as well as strong atmospheric expansions associated with
quasi-adiabatic cooling; see, e.g., \citet*{gai90} and \cite{kon92}
for theoretical work on the build-up of wave pressure due to shocks.
The impact of very strong shocks entailed that the time-averaged
temperatures are lowest in models with small tube spreading, i.e.,
high magnetic filling factors (e.g., case of $f = 0.1\%$) if
time-dependent ionization of hydrogen is taken into account.

\begin{figure}
\centering
\begin{tabular}{c}
\includegraphics[width=0.93\linewidth]{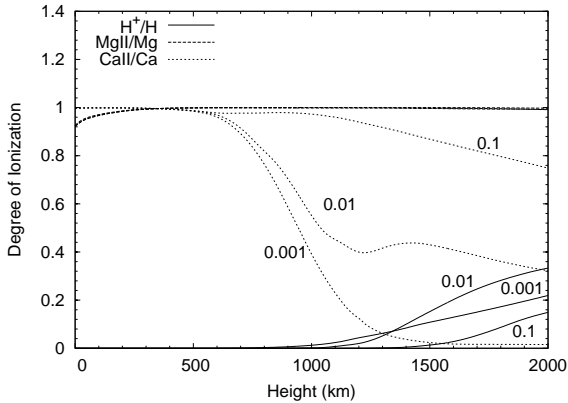}
\end{tabular}
\caption{
Comparison of the time-averaged ionization degrees for hydrogen,
magnesium and calcium concerning longitudinal wave computations with
time-dependent ionization and with consideration of frequency spectra.
We depict results for magnetic flux tubes with filling factors of
$f = 0.1\%$, 0.01\%, and 0.001\%.
\label{fig8}}
\end{figure}

\begin{figure}
\centering
\begin{tabular}{c}
\includegraphics[width=0.93\linewidth]{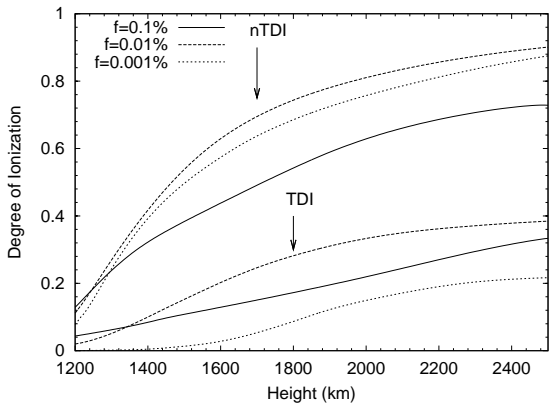}
\end{tabular}
\caption{
Comparison of time-averaged hydrogen ionization degrees for
longitudinal wave computations with (TDI) and without (nTDI)
time-dependent ionization regarding models based on
frequency spectra.  We depict results for magnetic flux tubes
with filling factors of $f = 0.1\%$, 0.01\%, and 0.001\%.
\label{fig9}}
\end{figure}

Finally, we studied the behaviour of the height-dependent time-averaged
hydrogen ionization in our set of models (see Figs.~8 and 9).
We found that the hydrogen ionization degrees steadily increase at
heights beyond 1200~km regardless of the magnetic filling factor, i.e.,
the tube top opening radius.  However, in general, they remain relatively
low (i.e., below 40\%) in models with time-dependent ionization (TDI),
whereas they are found to be relatively high (i.e., between 70\% and 90\%
at heights beyond 2000~km) in models without time-dependent ionization
(nTDI).  This is consistent with the general behaviour of time-dependent
hydrogen ionization as described, as in this case lower degrees of
ionization occur at most heights, including regions behind shocks,
including strong shocks owing to the time-delay of ionization in accord
to the time-dependent statistical rate equations (see Sect. 2.1).

We also evaluated time-dependent effects for the ionization of calcium
and magnesium (see Fig.~8), noting that Ca~II and Mg~II are of significant
importance for the facilitation of radiative cooling (see Sect.~2.2).  Our
simulations show that most of the magnesium is ionized to Mg~II at most heights
regardless of the geometrical filling factor of the respective flux tube.
However, for Ca~II the height-dependent behaviour of the ionization degree is
strongly impacted by the geometrical spread of the flux tubes (see Fig.~2).
Generally, for smaller values of $f$, associated with a lesser dilution of
the wave energy flux, a larger fraction of calcium is ionized to Ca~II.
The appropriate explanation is the lower dilution of the wave energy flux
in those models, leading to more effective shock heating and higher overall
temperatures.


\section{Summary and Conclusions}

We pursued sets of time-dependent model simulations for solar
magnetic flux tubes with focus on longitudinal tube waves with
and without the consideration of time-dependent ionization,
particularly pertaining to hydrogen.  The treatment of longitudinal
tube waves is motivated by a large variety of studies including
work by \cite{has08} arguing that longitudinal tube waves are able
to provide quasi-steady heating sufficient to explain the bright
solar network grains observed in Ca~II H and K.  We studied the dynamics
of flux tubes with different magnetic filling factors, i.e., 0.1\%,
0.01\%, and 0.001\%, corresponding to different tube top opening
radii, guided and motivated by previous observational results
\citep[e.g.,][]{spr83,sol96}.  It was found that the dynamic
tube structures are determined by a complex interplay between
shock heating, radiative and hydrodynamic cooling (with the latter
caused by wave pressure), dilution of the wave energy flux due
to the flux tube geometry, and time-dependent ionization.
We identified pronounced differences between the tubes models
especially in regard to tubes with different tube opening radii
and in response to the inclusion or omission of time-dependent
ionization.

Similar to the case of acoustic waves, time-dependent ionization
in LTW models leads to a large range of phenomena, such as over-
and underionization of the flow, increased temperature jumps at shocks,
and modified mean (i.e., time-averaged) temperature, density and
ionization structures.  The impact of time-dependent treatment of
ionization for flux tubes becomes evident through comparisons of
mean temperatures between the different types of models.  Regarding
monochromatic models it was found that time-dependent ionization
leads to lower average temperatures inside flux tubes compared to
time-independent ionization, an effect already known for acoustically
heated models.

However, the average temperatures inside of flux tubes
in models with and without time-dependent ionization are nonetheless
noticeably higher than in the acoustically heated external atmosphere.
If monochromatic longitudinal tube waves are used, the mean (i.e.,
time-averaged) temperatures of the tubes generally show a quasi-steady
increase with height.  However, if frequency spectra are used,
the mean temperatures, in essence, show no rise with height at all.
This behaviour is initiated by the formation of
very strong shocks (especially in narrow tubes, corresponding to
small magnetic filling factors) resulting in large-scale
quasi-adiabatic cooling, leading to unrealistically low temperatures.
This type of result has already previously been found in corresponding
1-D models of acoustic waves \cite[e.g.,][]{car92,car95,ramu03},
a behaviour deemed highly unrealistic \citep{ulm05}.

An appropriate enhancement of our current models will be the
calculation of self-consistent 3-D magnetohydrodynmic (MHD) models.
Those models face, however, the principal challenge of the necessity
to include both detailed multi-level 3-D radiative transfer and
3-D flows including the detailed treatment of shock formation
and shock interaction.  Important progress has already been
made \citep[e.g.,][]{stei09a,stei09b}, but further efforts
are needed to obtain of a concise picture.  Tentative insights
into the principal properties of these types of future models
can be attained through inspecting existing 3-D time-dependent
{\it non-magnetic} hydrodynamics models, which also consider
time-dependent non-equilibrium effects caused by hydrogen
\citep{lee06}, albeit various restrictive assumptions including
(but not limited to) the lack of back-coupling of the ionization
to the equation of state.  In this type of models it is found
that the build-up of strong shocks due to shock interaction is
largely absent, resulting in a lack of unrealistically high
cooling behind the strong shocks previously also referred to
as ``hydrodynamic refrigeration" \citep{cunm94}.  In this case
a quasi-steady rise of temperature with height is attained,
which appears to be in close resemblance to empirical solar
chromosphere models \citep[e.g.,][]{and89}.  Nonetheless,
our results based on time-dependent ionization and LTW
frequency spectra allow insight into the limiting case of
1-D geometry, while also noting that our models based on
time-dependent ionization and monochromatic waves are
expected to be approximately reflective of physical reality.

An alternative, or perhaps supplementary, way of supplying
chromospheric heating might be given through ambipolar diffusion
as described by \cite{kho12}.  Here the presence of neutrals,
together with the decrease with height of the collisional coupling,
leads to deviations from the classical magnetohydrodynamic behaviour
of the chromospheric plasma.  \cite{kho12} pointed out that a relative
net motion occurs between the neutral and ionized components, referred
to as ambipolar diffusion. According to this model, the dissipation of
currents in the chromosphere is enhanced by orders of magnitude due to
the action of ambipolar diffusion, as compared with the standard ohmic
diffusion.  The authors proposed that a significant amount of magnetic
energy can be released to the chromosphere just by existing force-free
10--40~G magnetic fields there.

Additional studies were given by \cite*{fed09}, \cite*{vig09}, and
\cite{erd10}.  \cite{fed09} studied the oscillatory response of the
3-D solar photosphere to the leakage of photospheric motion.  They found,
among other results, that high-frequency waves propagate from the
lower atmosphere across the transition region experiencing relatively
low reflection, and transmitting most of their energy into the corona,
and, furthermore, that the magnetic field acts as a waveguide for
both high- and low-frequency waves originating from the photosphere
and propagating up into the solar corona.  \cite{vig09} provided a
targeted study on wave propagation and energy transport in the
magnetic network of the Sun based on 2-D MHD simulations, which
among other results identified the limited capacity of acoustic
waves.  \cite{erd10} investigated the oscillatory modes
of a magnetically twisted compressible flux tube embedded
in a compressible magnetic setting, including applications
to solar magneto-seismology.

As an overarching statement concerning our study we conclude that
the significance of time-dependent ionization identified in the
simulations of longitudinal tube waves
is a stark motivation to also consider this type of effect in future
models of transverse and torsional tube waves.  This will
allow to obtain a more detailed picture of the dynamics and
energetics of solar-type outer atmospheres.  This obvious suggestion
is also supported by the repeatedly obtained finding that longitudinal
flux tube waves, as gauged through models considering time-dependent
ionization phenomena as done in the present study, are insufficent
to supply an adequte amount of energy for balancing coronal heating
in the view of early and updated estimates by \cite{gue07} and others.

\section*{Acknowledgments}
This work utilizes an OHD-MHD computer code package developed
by P. Ulmschneider and his group (which also included the authors of
this study) at the Institute for Theoretical Astrophysics, University
of Heidelberg, Germany, including subsequent augmentations.


{}

\end{document}